\documentclass[journal,twoside,web]{ieeecolor}
\usepackage{generic}

\usepackage{cite}
\usepackage{amsmath,amssymb,amsfonts}
\usepackage{graphicx}
\usepackage{textcomp}

\usepackage{subcaption}

\usepackage{xfrac}
\DeclareMathOperator*{\argmin}{argmin}

\usepackage{gensymb}

\usepackage[nolist]{acronym}

\usepackage{braket}

\usepackage{physics}

\usepackage{mathtools}

\usepackage{hyperref}

\usepackage{amsfonts}
\usepackage{nicefrac}

\usepackage[final]{pdfpages}

\usepackage{tabu}

\newtheorem{definition}{Definition}

\newtheorem{lemma}{Lemma}
\newtheorem{corollary}{Corollary}
\newtheorem{theorem}{Theorem}

\newtheorem{remark}{Remark}
\newtheorem{initialization}{Initialization}
\newtheorem{instantiation}{Instantiation}

\usepackage{algorithm} 
\usepackage{algpseudocode}

\providecommand{\openbox}{\leavevmode
  \hbox to.77778em{%
  \hfil\vrule
  \vbox to.675em{\hrule width.6em\vfil\hrule}%
  \vrule\hfil}}
\makeatletter
\DeclareRobustCommand{\qed}{%
  \ifmmode
    \eqno \def\@badmath{$$}
    \let\eqno\relax \let\leqno\relax \let\veqno\relax
    \hbox{\openbox}%
  \else
    \leavevmode\unskip\penalty9999 \hbox{}\nobreak\hfill
    \quad\hbox{\openbox}%
  \fi
}
\makeatother

\def\BibTeX{{\rm B\kern-.05em{\sc i\kern-.025em b}\kern-.08em
    T\kern-.1667em\lower.7ex\hbox{E}\kern-.125emX}}
\newcommand{\mt}{m_{\mathcal{T}}}
\newcommand{\Et}{\mathbb{E}_\mathcal{T}}
\newcommand{\T}{\mathcal{T}}
\newcommand{\X}{\mathcal{X}}
\newcommand{\U}{\mathcal{U}}
\newcommand{\IntSet}{\mathbb{Z}}
\newcommand{\R}{\mathbb{R}}
\newcommand{\I}{\mathbb{I}}

\newcommand{\A}{\mathcal{A}}
\newcommand{\del}{\partial}

\begin{document}

\title{Systematic, Lyapunov-Based, Safe and Stabilizing Controller Synthesis for Constrained Nonlinear Systems}

\author{Reza Lavaei, \IEEEmembership{Graduate Student Member, IEEE}, Leila Bridgeman, \IEEEmembership{Member, IEEE}
\thanks{}
\thanks{Reza Lavaei and Leila Bridgeman are with the Department of Mechanical Engineering and Materials Science at Duke University, Durham NC, USA (email:  reza.lavaei@duke.edu; leila.bridgeman@duke.edu), corresponding author: Reza Lavaei}
}

\maketitle

\begin{abstract}
A controller synthesis method for state- and input-constrained nonlinear systems is presented that seeks continuous piecewise affine (CPA) Lyapunov-like functions and controllers simultaneously. Non-convex optimization problems are formulated on triangulated subsets of the admissible states that can be refined to meet primary control objectives, such as stability and safety, alongside secondary performance objectives. A multi-stage design is also given that enlarges the region of attraction (ROA) sequentially while allowing exclusive performance for each stage. A clear boundary for an invariant subset of closed-loop system's ROA is obtained from the resulting Lipschitz Lyapunov function. For control-affine nonlinear systems, the non-convex problem is formulated as a series of conservative, but well-posed, semi-definite programs. These decrease the cost function iteratively until the design objectives are met. Since the resulting CPA Lyapunov-like functions are also Lipschitz control (or barrier) Lyapunov functions, they can be used in online quadratic programming to find minimum-norm control inputs. Numerical examples are provided to demonstrate the effectiveness of the method.
\end{abstract}


\section{Introduction} \label{sec:introduction}
\IEEEPARstart{L}{y}apunov theory has been instrumental to  stable \cite{lyapRev1,lyapRev2,lyapRev3,lyapRev4,lyapRev5,lyapRev6,lyapRev7,lyapRev8} and safe \cite{CBF2007,barrier2015,barrier2016,CBF2016,Ames2016,barrier2020,barrier2020Conf} control design. While for linear systems it provides straightforward stability criteria for analysis and design, since the existence of a Lyapunov function can be assured or denied simply by solving a set or linear matrix inequalities (LMIs), there is no systematic means to ensure Lyapunov stability for general, nonlinear systems. When physical limitations or operating considerations constrain the state and inputs, the problems of stability and feasibility must be tackled in tandem, further complicating analysis. Ignoring constraints at best results in unexpected
closed loop behaviour if the constraints are activated by physical boundaries, and at worst is hazardous if the system operates in unsafe regions. By limiting the choice of controller and Lyapunov functions to a particular class, this paper formulates a controller design method for state- and input-constrained nonlinear systems as an offline optimization problem on a triangulated subset of the admissible states.


Lyapunov stability of a closed loop system is verified by existence of a Lyapunov function. Safety (respecting state and input constraints), is ensured if the system is initialized in a sublevel set of the Lyapunov function which is itself a subset of the admissible states where the control constraints are also respected. By finding an explicit Lyapunov function alongside the controller, not only can the boundary of the \ac{ROA} and safe initial sets be specified, but also the convergence rate can be characterized. Moreover, it can eliminate non-trivial \textit{a priori} design choices such as invariant sets and control Lyapunov functions that are computationally taxing. However, in most design methods for state- and input-constrained nonlinear systems, the Lyapunov function is found either after controller design, or characterized \textit{a priori} \cite{nonMPC1998,nonMPCExTube2006,nonMPCexplicit2012,ExpMPCLike2017,nonMPC-Lyap2005,nonMPC-Lyap2006,Ames2016,CBF2007,CBF2016,CBF2018}. For instance, \ac{MPC} implies existence of a Lyapunov function by appropriate choices of terminal `ingredients' \cite{nonMPC1998} like the terminal set, terminal cost, and terminal stabilizing controller. Finding an explicit Lyapunov function and the controller can only be  done afterwards via taxing explicit nonlinear MPC \cite{nonMPCExTube2006,nonMPCexplicit2012,ExpMPCLike2017}. Also, Lyapunov-based methods rely on finding a Lyapunov-like function such as \ac{CLF} or \ac{CBF} first \cite{nonMPC-Lyap2005,nonMPC-Lyap2006,Ames2016,CBF2007,CBF2016,CBF2018}, and even with one at hand, the controller is usually realized online using \ac{QP}, which is not desirable when computing resources are limited.


This paper's inspiration comes from a versatile stability analysis method given in \cite{gieslRevCPA2013}, which formulates the search for a \ac{CPA} Lyapunov function as a linear program defined on a triangulated subset of the \ac{ROA} of an exponentially stable equilibrium point. Since the linear program may not be feasible, using a standard triangulation, a refinement algorithm was designed to obtain a \ac{CPA} Lyapunov function in a finite number of steps \cite{gieslRevCPA2013}. Attempting to turn it into a design method, \cite{Doban} assumed that a \ac{CLF} and the gradient of a CPA controller were known \textit{a priori}. Aside from difficulty in finding a \ac{CLF}, assigning gradients directly influences the feasibility of the proposed linear program. Here, we build upon these  ideas to develop offline optimization-based design methods that ensure stability and safety while eliminating the need for unclear \textit{a priori} design choices like the \ac{CLF}s and controller gradients in \cite{Doban}, and \ac{CBF}s, and terminal invariant sets often associated with \ac{MPC}. Moreover, unlike design by sum-of-squares \cite{sos2004,sos2009,sos2010}, the method is not limited to polynomial systems.



Preliminary results employing uniform triangulation refinements and focusing on exponential stability appeared in \cite{me}. Here, these results are extended to also confront safety by iteratively increasing the \ac{ROA}. Moreover, the modified techniques admit varied refinement schemes, and improve secondary performance objectives. For control-affine systems, the proposed offline non-convex optimizations are formulated as iterative \acp{SDP} to alleviate computations in finding a controller, which can then be used as a feasible initial point for further improvement in the non-convex optimizations. Combined with proposed triangulation refinements, the CPA controller and Lyapunov functions become flexible tools for control design, in particular for control-affine nonlinear constrained systems, where the proposed iterative SDP method is applicable. Numerical examples demonstrate the effectiveness of the method.



\section{Preliminaries} \label{sc:Prelims}
\textbf{Notation.} The interior, boundary, and closure of $\Omega\in\R^n$ are denoted by $\Omega\degree$, $\del\Omega$, and $\bar{\Omega}$, respectively. The set of real-valued functions with $r$ times continuously differentiable partial derivatives over their domain is denoted by $\mathbb{C}^r$. The $i^{\textrm{th}}$ element of a vector $x$ is denoted by $x^{(i)}$. The element in the $i^{\textrm{th}}$ row and $j^{\textrm{th}}$ column of a matrix $G$ is denoted by $G^{(i,j)}$. The preimage of a function $f$ with respect to a subset $\Omega$ of its codomain is defined by $f^{-1}(\Omega)=\{x\mid f(x) \in \Omega \}$. The transpose and Euclidean norm of $x\in\R^n$ are denoted by $x^\intercal$ and $||x||$, respectively. The set of all compact subsets $\Omega\subset\R^n$ satisfying i) $\Omega\degree$ is connected and contains the origin, and ii) $\Omega=\overline{\Omega\degree}$, is denoted by $\mathfrak{R}^n$. The vector of ones in $\R^n$ is denoted by $1_n$. 

For a constrained autonomous system, safety can be ensured in a positive-invariant subset of the feasible region.

\begin{definition}[Positive-invariance {\cite[Ch\;11]{BorrelliBook}}] \label{def:PI}
Consider the system $\dot{x} = g(x)$, $x\in\X\subset\R^n$, where $g(\cdot)$ is a Lipschitz map, and $\X$ is compact. A set $\A\subseteq\X$ is positive-invariant if $x(t_0)\in\A$ implies $x(t)\in\A$ for all $t>t_0$.
\end{definition}

When the input is also constrained, control-invariance, defined next, ensures safety.

\begin{definition}[Control-invariance {\cite[Ch\;11]{BorrelliBook}}]
Consider the system $\dot{x} = g(x,u)$, $x\in\X\subset\R^n$, $u\in\U\subset\R^m$, where $g(\cdot)$ is a Lipschitz map, and $\X,\U$ are compact. A set $\A\subseteq\X$ is control-invariant if there exists $u^\ast:\A\rightarrow\U$ that makes $\A$ positive-invariant for $\dot{x}=g(x,u^\ast(x))$.
\end{definition}

A \ac{ROA} is a positive-invariant set in which $x(t)\rightarrow0$ as $t\rightarrow +\infty$. In this paper, sublevel sets of Lipschitz Lyapunov-like functions are used to find safe set and/or \ac{ROA}s. These functions will be constructed on a triangulated subset of $\R^n$. Their sublevel sets provide positive-invariant sets and/or \ac{ROA}s. The required definitions are given next.

\begin{definition}[Affine independence{\cite{gieslRevCPA2013}}] \label{def:affDepVecs}
A collection of vectors $\{x_0,\ldots,x_n\}$ in $\R^n$ is called affinely independent if $x_1-x_0,\ldots,x_n-x_0$ are linearly independent. \qed
\end{definition}

\begin{definition}[$n$-simplex {\cite{gieslRevCPA2013}}] \label{def:simplex}
An $n$-simplex is the convex combination of $n+1$ affinely independent vectors in $\R^n$, denoted $\sigma{=}\textrm{co}(\{x_j\}_{j=0}^n)$, where $x_j$'s are called vertices. \qed
\end{definition}

\noindent In this paper, simplex always refers to $n$-simplex. By abuse of notation, $\T$ will refer to both a collection of simplexes and the set of points in all the simplexes of the collection. 

\begin{definition} [Triangulation {\cite{gieslRevCPA2013}}] \label{def:triangulation}
A set $\T\in\mathfrak{R}^n$ is called a triangulation if it is a finite collection of $\mt$ simplexes, denoted $\T=\{\sigma_i\}_{i=1}^{\mt}$, and the intersection of any of the two simplexes in $\T$ is either a face or the empty set. 

The following two conventions are used throughout this paper for triangulations and their simplexes. Let $\T=\{\sigma_i\}_{i=1}^n$. Further, let $\{x_{i,j}\}_{j=0}^n$ be $\sigma_i$'s vertices, making $\sigma_i=\textrm{co}(\{x_{i,j}\}_{j=0}^n)$. The choice of $x_{i,0}$ in $\sigma_i$ is arbitrary unless $0\in\sigma_i$, in which case $x_{i,0}=0$. The vertices of the triangulation $\T$ that are in $\Omega\subseteq\T$ is denoted by $\mathbb{E}_\Omega$. \qed
\end{definition}

\begin{definition} [Triangulable Set] \label{def:trinagulableSets}
A compact, connected subset of $\R^n$ that has no isolated points, and can be exactly covered by a finite number of simplexes is called triagulable. \qed
\end{definition}

\noindent As an example, full-dimensional polytopes \cite[Ch\;5]{BorrelliBook} in $\R^n$ are triangulable.  

\begin{definition} [Constraint Surfaces of a Triangulation]
Let $\T$ be the triangulation of a trinagulable set $\Omega\in\R^n$. The surface $\mathcal{H}\subset\T$ is called a constraint surface in $\T$ if it is exactly covered by the faces of some simplexes in $\T$. \qed
\end{definition} 

\begin{definition} [CPA interpolation \cite{gieslRevCPA2013}] \label{def:CPAfunction}
Consider a triangulation $\T{=}\{\sigma_i\}_{i=1}^{\mt}$, and a set $\mathbf{W}{=}\left\{ W_x \right\}_{ x\in \Et } {\subset} \R$. The unique, CPA interpolation of $\textbf{W}$ on $\T$, denoted $W:\T{\rightarrow}\R$, is affine on each $\sigma_i{\in}\T$ and satisfies $W(x){=}W_x$, $\forall x{\in}\Et$. \qed
\end{definition}

\begin{remark}[{\!\!\cite[Rem.\;9]{gieslRevCPA2013}}] \label{rem:nablaLinear}
    Given $\T=\{\sigma_i\}_{i=1}^{\mt}$ and $\mathbf{W}$, the CPA interpolation assigns a unique affine function $W(x)=x^\intercal\nabla{W}_i+\omega_i$ to each $\sigma_i\in\T$. The $\nabla{W}_i$ is linear in the elements of $\mathbf{W}$ and can be computed as follows. Let $\sigma_i=\textrm{co}(\{x_{i,j}\}_{j=0}^n)$, and $X_i\in\R^{n\times n}$ be a matrix that has $x_{i,j}-x_{i,0}$ as its $j$-th row. Since the elements of $\{x_{i,j}\}_{j=0}^n$ are affinely independent, $X_i$ is invertible. Each $x_{i,j}$ is an element of $\Et$, so it has a corresponding element in $\mathbf{W}$, denote $W_{x_{i,j}}$. Let $\bar{W}_i\in\R^n$ be a vector that has $W_{x_{i,j}}-W_{x_{i,0}}$ as its $j$-th element. Then, ${\nabla W}_i = X^{-1}_i \bar{W}_i$. \qed
\end{remark}

The following theorem from \cite{gieslRevCPA2013} bounds the time derivative of a CPA function above on a simplex using its values at the vertices of that simplex using Taylor's theorem. 

\begin{lemma}[{\!\!\cite{gieslRevCPA2013}}] \label{lem:gieseldWBound}
Consider the system
\begin{equation} \label{eq:AutSystem}
 \dot{x} = g(x), \;\; x\in \X \in \mathfrak{R}^n, 
 \end{equation}
where $g:\R^n\rightarrow\R^n$ is in $\mathbb{C}^2$. Let $\T=\{\sigma_i\}_{i=1}^{\mt} \subseteq \X$ be a triangulation, and $V:\T\rightarrow\R$ be the CPA interpolation of a set $\mathbf{V}=\{V_x\}_{x\in{\Et}}$. Consider a point $x\in\T\degree$. Let $D^+V(x) = \textrm{lim\,sup}_{h\rightarrow0^+}\sfrac{(V(x+hg(x))-V(x))}{h}$ be the Dini derivative of $V$ at $x$, which equals $\dot{V}(x)$ where $V\in\mathbb{C}^1$. For an arbitrary $x\in\T\degree$, there exists a $\sigma_i=\{x_{i,j}\}_{j=0}^n\in\T$ so that for small enough $h>0$, $\textrm{co}(x,x+hg(x))\subset\sigma_i$. Let $0\leq\alpha_j\leq 1$, where $j\in\IntSet_0^n$ and $\sum_{j=0}^n\alpha_j=1$, be the unique set of coefficients satisfying $x=\sum_{j=0}^n\alpha_ix_{i,j}$. Then
\begin{equation} \label{eq:gieselInequality}
    D^+V(x) \leq \sum_{j=0}^n \alpha_j \left( g(x_{i,j})^\intercal\nabla{V}_i + c_{i,j}\beta_i 1_n^\intercal l_i \right),
\end{equation}
where $l_i\in\mathbb{R}^n$ satisfies $l_i\geq|\nabla{V}_i|$, and 
\begin{flalign} 
    &\beta_i \geq  \max_{p,q,r\in\IntSet_1^n} \max_{\xi\in\sigma_i} \left| \left. \sfrac{\del^2 g^{(p)}}{\del x^{(q)}\del x^{(r)}} \right|_{x=\xi} \right|, \textrm{ and} \label{eq:beta} \\
    &c_{i,j}{=}\frac{n}{2} ||x_{i,j} {-} x_{i,0}|| (\max_{k\in\IntSet_1^n} ||x_{i,k}{-}x_{i,0}|| {+} ||x_{i,j}{-}x_{i,0}||). \nonumber
\end{flalign}
\qed
\end{lemma}

Note that in \eqref{eq:beta}, $\beta_i$ bounds the largest absolute value of the elements of the Hessian of $g(x)$ on $\sigma_i$ above.

\section{Controller Characterization} \label{sc:ControlCharacter}
The goal is to turn the analysis method of \cite{gieslRevCPA2013} into a design method for state- and input-constrained control systems by finding a state-feedback controller. While an optimization problem could simply be derived by directly applying \cite[Thm\;1]{gieslRevCPA2013} to the closed-loop of a plant and a parametrized controller structure, this does not readily lead to a well-posed, convex optimization problem and a synthesis method. Consequently, the theorems that follow parallel those of \cite[Thm\;1]{gieslRevCPA2013} with appropriate modifications to establish a decrease in a closed-loop Lyapunov function. These changes are critical to the proposed iterative synthesis method of Section\;\ref{sc:IterativeAlgos}. First, piecewise twice continuous differentiability on a triangulation is defined. 
\begin{definition} \label{def:C2functionsPiecewise}
A continuous function $g(x)\in\R^n$ is piecewise in $\mathbb{C}^2$ on a triangulation $\T=\{\sigma_i\}_{i=1}^{\mt}$, denoted $g\in\mathbb{C}^2(\T)$, if it is in $\mathbb{C}^2$ on $\sigma_i$ for all $i\in\IntSet_1^{\mt}$. \qed
\end{definition}

When taking derivatives, if $\xi$ is on the common face of some simplexes, the surrounding notation will clarify which one should be considered. When considering $\xi\in\sigma_i$, the related limits should be evaluated in directions, $y\in R^n$ where $\xi+hy\in\sigma_i$ as $h \rightarrow 0$.


\begin{theorem} \label{thm:genOpt}
Consider the system
\begin{align} \label{eq:controlSystem}
    \dot{x} = g(x,u), \; x\in\X\in\mathfrak{R}^n, \; u\in\U\in\mathfrak{R}^m, \; g(0,0)=0.
\end{align}
Given a triangulable set $\Omega\subseteq\X$, let $\T =\{\sigma_i\}_{i=1}^{\mt}$ be its triangulation. Suppose that a class of Lipschitz controllers $\mathcal{F}=\{u(\cdot,\boldsymbol{\lambda})\}$ parameterized by $\boldsymbol{\lambda}$ has at least one element and satisfies $u(0,\lambda)=0$, $u(\cdot,\boldsymbol{\lambda})\in\mathbb{C}^2(\T)$, and $g_{\boldsymbol{\lambda}}(\cdot) \coloneqq g(\cdot,u(\cdot,\boldsymbol{\lambda}))\in\mathbb{C}^2(\T)$ is Lipschitz on $\T$, and $u(x,\boldsymbol{\lambda})\in\U$ for $\forall x\in\Et$ implies $u(x,\boldsymbol{\lambda})\in\U$ for $\forall x\in\T$. Consider the following nonlinear program.
\begin{subequations} \label{eq:genOpt} 
    \begin{alignat}{2}
        [\mathbf{V}^\ast,\; &\mathbf{L}^\ast,\; \boldsymbol{\lambda}^\ast,\; a^\ast,\; \mathbf{b}^\ast] = && \argmin_{\mathbf{V},\; \mathbf{L},\; \boldsymbol{\lambda},\; a,\; \mathbf{b}} \;\; J(\mathbf{V}, \mathbf{L}, \boldsymbol{\lambda}, a, \mathbf{b})  \nonumber \\
        \textrm{s.t.} \;\; & a\geq1,\; b_1 > 0, && \label{eq:V0b1Constraint} \\
        & b_1||x||^a \leq V_x, &&\forall x\in\Et, \label{eq:myVconstraint} \\
        & |{\nabla V}_i| \leq l_i, &&\forall i\in\IntSet_1^{\mt}, \label{eq:nablaConstraint} \\
        & u(x_{i,j},\boldsymbol{\lambda})\in\U, &&\forall i\in\IntSet_1^{\mt}, \; \forall j\in\IntSet_0^n, \label{eq:uConstraintGeneral} \\
        & D^+_{i,j}V \leq -b_2 V_{x_{i,j}}, \;\; &&\forall i\in\IntSet_1^{\mt}, \; \forall j\in\IntSet_0^n, \label{eq:myDv}
        \end{alignat}
\end{subequations}

\noindent where $D^+_{i,j}V=g_{\boldsymbol{\lambda}}(x_{i,j})^\intercal {\nabla V}_i + c_{i,j}\beta_i 1_n^\intercal l_i$, and $\mathbf{V}=\{V_x\}_{x\in\Et}\subset\R$ and $\mathbf{L}=\{l_i\}_{i=1}^{\mt}\subset\R^n$, and $\mathbf{b}=\{b_1,b_2\}\subset\R$, and $J(\cdot)$ is a cost function, and for $u(\cdot,\boldsymbol{\lambda})$ satisfying \eqref{eq:uConstraintGeneral},
\begin{flalign}\label{eq:betaAndc}
    &\beta_i \geq \max_{p,q,r\in\IntSet_1^n} \max_{\xi\in\sigma_i} \left| \left. \sfrac{\del^2 g^{(p)}_{\boldsymbol{\lambda}}}{\del x^{(q)}\del x^{(r)}} \right|_{x=\xi} \right|, \textrm{ and} & \\
    &c_{i,j}{=} \frac{n}{2} ||x_{i,j} {-} x_{i,0}|| (\max_{k\in\IntSet_1^n} ||x_{i,k}{-}x_{i,0}|| {+} ||x_{i,j}{-}x_{i,0}||).& \nonumber
\end{flalign}

\noindent The optimization \eqref{eq:genOpt} is feasible, and the CPA function $V^\ast:\T\rightarrow\R$, constructed from $\textbf{V}^\ast$, satisfies $b_1^\ast||x||^{a^\ast} \leq V^\ast(x)$ and $D^+V^\ast(x)\leq b_2^\ast V^\ast(x)$ for all $x\in\T\degree$. \qed
\end{theorem}

\begin{proof}
To see that \eqref{eq:genOpt} is feasible, note that $V_x=b_1||x||^a$ with any $a\geq1$ and $b_1>0$ satisfies \eqref{eq:V0b1Constraint}--\eqref{eq:myVconstraint} and can be used to compute a feasible solution $l_i=|\nabla{V}_i|$ for \eqref{eq:nablaConstraint} using Remark\;\ref{rem:nablaLinear}. By assumption, a feasible $\boldsymbol{\lambda}$ exists satisfying \eqref{eq:uConstraintGeneral}. Using these feasible values, finite $\beta_i$ satisfying \eqref{eq:betaAndc} can be chosen and $g_{\boldsymbol{\lambda}}(\cdot)$ is always finite because $g_{\boldsymbol{\lambda}}(\cdot)\in\mathbb{C}^2(\T)$. Likewise, $c_{i,j}$ is finite because each $\sigma_i$ is compact, making the left-hand side of \eqref{eq:myDv} finite for each $i\in\IntSet_1^{\mt}$ and $j\in\IntSet_0^n$. Note that if $x_{i,j}=0$, then $g_{\boldsymbol{\lambda}}(x_{i,j})=0$ and by convention, $j=0$, so $c_{i,j}=0$, making $D^+_{i,j}V=0$, making any $b_2$ feasible. Thus, there exists $b_2{\in}\R$ that satisfies \eqref{eq:myDv} for all $i{\in}\IntSet_1^{\mt}$ and $j{\in}\IntSet_0^n$. 

The remainder of the proof is devoted to showing that for the closed-loop system, $\dot{x}=g_{\boldsymbol{\lambda}}(x)$, the solution $V^\ast(x)$ satisfies $b_1^\ast||x||^{a^\ast}\leq V^\ast(x)$ and $D^+V^\ast(x)\leq b_2^\ast V^\ast(x)$ for all $x\in\T\degree$. By assumption,  \eqref{eq:uConstraintGeneral} implies $u(x,\boldsymbol{\lambda})\in\U$ for all $x\in\T$. For simplicity, asterisks are dropped. Constraints \eqref{eq:V0b1Constraint}--\eqref{eq:myVconstraint} ensure that $b_1 ||x||^a \leq V(x)$ for all $x\in\T$ since $V:\T\rightarrow\R$ is a CPA function. It remains to show that \eqref{eq:nablaConstraint} and \eqref{eq:myDv} verify $D^+V(x)\leq -b_2V(x)$ for all $x\in\T\degree$. For simplicity, let $g(x)=g(x,u(\boldsymbol{\lambda},x))$. The assumptions of Lemma\;\ref{lem:gieseldWBound} are verified by \eqref{eq:nablaConstraint},\eqref{eq:betaAndc}. Applying \eqref{eq:gieselInequality}, \eqref{eq:myDv}, and the fact that $V(x)\geq0$ is affine on each $\sigma_i$ shows that $D^+V(x) \leq \sum_{j=0}^n \alpha_j D^+_{i,j} V \leq -b_2 \sum_{j=0}^n \alpha_j V_{x_{i,j}} = -b_2 V(x)$, where $x=\sum_{j=0}^n\alpha_jx_{i,j}\in\T\degree$ is arbitrary, and $\{\alpha_j\}_{j=0}^n\subset\R$ is the unique corresponding set satisfying $\sum_{j=0}^n \alpha_j=1$ and $0\leq\alpha_j\leq1$. Like \cite{Doban}, as a relaxation of Theorem\;\ref{lem:gieseldWBound}, it is assumed that $g_{\boldsymbol{\lambda}}(\cdot)\in\mathbb{C}^2(\T)$, not everywhere. Since $x\in\T\degree$ was an arbitrary point, $D^+V(x)\leq -b_2V(x)$ for all $x\in\T\degree$ is verified.
\end{proof}

As will be discussed later, formulating Lyapounov-like functions depends on having $b_2^\ast>0$ in \eqref{eq:genOpt}. However, even if $b_2^\ast \leq 0$ is found, a connected subset of $\T$ in which $D_{i,j}^+V^\ast$ is positive might exist. This is described by the following corollary.

\begin{corollary} \label{cor:mightFindAPosb2}
Suppose that $b_2^\ast\leq0$ is found in Theorem\;\ref{thm:genOpt}. Let $\mathbb{I}= \{i\in\IntSet_1^{\mt} \mid D^+V^\ast_{x_{i,j}} < 0, \forall j\in\IntSet_0^n, x_{i,j}\neq0 \}$, and $\hat{\mathcal{T}}=\{\sigma_i\}_{i\in\mathbb{I}}$. Then, $V^\ast(x)$ satisfies $b_1^\ast||x||^{a^\ast}\leq V^\ast(x)$ and $D^+V^\ast(x)\leq \hat{b}_2^\ast V^\ast(x)$ for all $x\in\hat{\T}\degree$, where $\hat{b}_2^\ast \coloneqq \min \Set{-D^+_{i,j}V^\ast / V^\ast_{x_{i,j}} \mid i\in\mathbb{I}, j\in\IntSet_j^n, x_{i,j} \neq 0 }$. \qed
\end{corollary}

\begin{proof}
For all simplexes in $\mathcal{E}$, $D_{i,j}^+V^\ast$ is negative except at $0$ (if $0$ is in the triangulation) where it is zero, making $\hat{b}_2^\ast$ positive. Since $\hat{\T}$ is a triangulation and is a subset of the initial $\T$ on which $b_2^\ast\leq 0$ was found, Theorem\;\ref{eq:genOpt} holds for $\T\coloneqq\hat{\T}$ and $b_2^\ast \coloneqq \hat{b}_2^\ast$.
\end{proof}

Even if the right-hand side of \eqref{eq:myDv} is replaced by $-b_2$, after solving \eqref{eq:genOpt}, $D^+V^\ast(x)\leq -\tilde{b}_2^\ast V^\ast(x)$ can be written with $\tilde{b}_2^\ast$ obtained as follows.

\begin{corollary} \label{cor:simpleRHSforDV}
If the right-hand side of \eqref{eq:myDv} is replaced by $-b_2$, then the claims of Theorem\;\ref{thm:genOpt} hold for $b_1^\ast||x||^{a^\ast} \leq V^\ast(x)$, and $D^+V^\ast(x)\leq -\tilde{b}_2^\ast V^\ast(x)$ for all $x\in\T\degree$, where $\tilde{b}_2^\ast= \min \Set{b_2^\ast/V_{i,j}^\ast  \mid i\in\IntSet_{1}^{\mt},j\in\IntSet_0^n, x_{i,j}\neq 0}$.
\end{corollary}

\begin{proof}
Applying \eqref{eq:gieselInequality}, the modified \eqref{eq:myDv} with $-b_2$ as the right-hand side, and the fact that $V^\ast(x)>0$ (when it is zero, any $b_2$ works) is affine on each $\sigma_i$ shows that $D^+V^\ast(x) \leq \sum_{j=0}^n \alpha_j D^+_{i,j} V^\ast \leq - \sum_{j=0}^n \alpha_j (b_2/V_{x_{i,j}}^\ast) V_{x_{i,j}}^\ast \leq -\sum_{j=0}^n \alpha_j \tilde{b}_2^\ast V_{x_{i,j}}^\ast \leq - \tilde{b}_2^\ast \sum_{j=0}^n \alpha_j V_{x_{i,j}}^\ast  = -\tilde{b}_2^\ast V(x)^\ast$, where $x=\sum_{j=0}^n\alpha_j x_{i,j}\in\T\degree$ is arbitrary, and $\{\alpha_j\}_{j=0}^n\subset\R$ is the unique corresponding set satisfying $\sum_{j=0}^n \alpha_j=1$, and $0\leq\alpha_j\leq1$. Since $x\in\T\degree$ was an arbitrary point, $D^+V^\ast(x)\leq -\tilde{b}_2^\ast V^\ast(x)$ for all $x\in\T\degree$ is verified.
\end{proof}

In the following theorems, the right-hand side of \eqref{eq:myDv} remains as-is to impose a desired upper-bound on the decay rate of the state norm. However, Corollary\;\ref{cor:simpleRHSforDV} allows replacing it with $-b_2$ if needed, in which case the decay rate can be found once a solution for \eqref{eq:genOpt} is obtained.

\subsection{CPA Controller for Control-affine Systems}
In practice, it may not be obvious how to apply Theorem\;\ref{thm:genOpt} for control design. For one, finding a control structure in which point-wise feasibility on vertices of a triangulation implies feasibility at all points in the triangulation is not trivial. Once the control structure is chosen, its first and second derivatives may need to be constrained to compute $\beta_i$ in \eqref{eq:betaAndc}. Moreover, as will be discussed later, $b_2$ needs to be positive to ensure desired objectives such as Lyapunov stability. A more practical characterization of \eqref{eq:genOpt} for control-affine systems with polytopic input constraints is given next.

\begin{instantiation} \label{insta:semiProg}
Consider the constrained control system
\begin{align} \label{eq:controlAffineSystem}
    \dot{x} = f(x) + G(x)u, \;\;x\in\X\in\mathfrak{R}^n, \;\;u\in\U\in\mathfrak{R}^m,
\end{align}
where $\U=\Set{u\in\R^m \mid H u \leq h_c}$. Given a triangulable set $\Omega\subseteq\X$ and its triangulation, $\T =\{\sigma_i\}_{i=1}^{\mt}$, suppose that both $f(\cdot),G(\cdot)\in\mathbb{C}^2(\T)$. Let $u$ be CPA on $\T$, i.e. $u:\T \rightarrow \R^m$, where  ${u^{(s)}}_i = x^\intercal \nabla{u^{(s)}}_i+\omega^{(s)}_i$, $\forall s\in\IntSet_1^m, \forall i\in\IntSet_1^{\mt}$. Let $\mathbf{y} =  [\mathbf{V}, \mathbf{L}, \mathbf{U}, \mathbf{Z}, a, \mathbf{b}]$ be the unknowns, where $\mathbf{V}=\{V_x\}_{x\in\Et}\subset\R^n$ and $\mathbf{L}=\{l_i\}_{i=1}^{\mt}\subset\R^n$, and $\mathbf{U}=\{u_x\}_{x\in\Et}\subset\R^m$, and $\mathbf{Z}=\{z_i\}_{i=1}^{\mt}\subset\R$, and $a\in\R$, and  $\mathbf{b}=\{b_1,b_2\}\subset\R$. The following optimization is an instantiation of \eqref{eq:genOpt}.
\begin{subequations} \label{eq:SemiProg} 
    \begin{alignat}{2}
        \mathbf{y}^\ast &= \argmin_{\mathbf{y}} \;\; J(\mathbf{y})  \nonumber \\
        \textrm{s.t.} \;\; & a\geq 1,\; b_1 > 0, && \label{eq:SemiV0b1Constraint} \\
        & b_1||x||^a \leq V_x, && \forall x\in\Et, \label{eq:SemiVconstraint} \\
        & |{\nabla V}_i| \leq l_i, && \forall i\in\IntSet_1^{\mt}, \label{eq:SemiNablaConstraint} \\
        & H u_x \leq h_c, && \forall x \in \Et\backslash\{0\}, \label{eq:SemUconstraint} \\
        & |\nabla{u^{(s)}}_i| \leq z_i, && \forall i\in\IntSet_1^{\mt}, \; \forall s\in\IntSet_1^m, \label{eq:SemDuConstraint} \\
        & D^+_{i,j}V \leq -b_2 V_{x_{i,j}}, \quad && \forall i\in\IntSet_1^{\mt}, \; \forall j\in\IntSet_0^n,  \label{eq:semDv}
        \end{alignat}
\end{subequations}

\noindent where $D^+_{i,j}V=\phi_{i,j} + u_{x_{i,j}}^\intercal G(x_{i,j})^\intercal\nabla{V}_i + c_{i,j}\eta_iz_i1_n^\intercal l_i$ and $\phi_{i,j} = f(x_{i,j})^\intercal {\nabla V}_i {+} c_{i,j}\mu_i 1_n^\intercal l_i$, and $J(\cdot)$ is a cost function, and $c_{i,j}$ is given in \eqref{eq:betaAndc}, and
\begin{flalign}\label{eq:affineB}
    &\mu_i{=}\max_{p,q,r\in\IntSet_1^n} \max_{\xi\in\sigma_i} \left| \left. \sfrac{\del^2 f^{(p)}}{\del x^{(q)}\del x^{(r)}}\right|_{x=\xi}\right| + \ldots & \\ 
    & \sum_{s=1}^m \left| \left. \sfrac{\del^2 G^{(p,s)}}{\del x^{(q)}\del x^{(r)}} \right|_{x=\xi} \right| \max_{u^{(s)}} \left|\textrm{proj}_s(\U)\right|, \textrm{ and } & \nonumber \\
    &\eta_i{=}\!\max_{p,q,r{\in}\IntSet_1^n} \max_{\xi{\in}\sigma_i}\!\sum_{s{=}1}^m \!\left| \left. \sfrac{\del G^{(p,s)}}{\del x^{(q)}}\right|_{x{=}\xi} \right|\! {+} \!\left| \left. \sfrac{\del G^{(p,s)}}{\del x^{(r)}}\right|_{x{=}\xi} \right|, \nonumber &
\end{flalign}
\noindent where $\textrm{proj}_s(\U)$ projects $\U$ onto the $s$-th axis of $\R^m$. \qed
\end{instantiation}

\begin{corollary} \label{cor:instantiation}
The optimization \eqref{eq:SemiProg} is feasible, and the CPA function $V^\ast$ satisfies $b_1^\ast||x||^{a^\ast} \leq V^\ast(x)$ and $D^+V^\ast(x)\leq b_2^\ast V^\ast(x)$ for all $x\in\T\degree$. \qed
\end{corollary}

\begin{proof}
Consider any $\sigma_i{\in}\T$. By generalizing \cite[Lem\;III.1]{Doban} to multi-input systems with polytopic input constraints, the right-hand side of \eqref{eq:betaAndc} can be bounded above by $\mu_i+\eta_i z_i$, where $z_i\geq |\nabla{u^{(s)}}_i|$, using the Triangle Inequality. Considering $z_i$ as an optimization variable, and replacing $\beta_i$ with $\mu_i{+}\eta_i z_i$, and including $z_i{\geq} |\nabla{u^{(s)}}_i|$ in \eqref{eq:genOpt}, \eqref{eq:SemiProg} is obtained. Moreover, since $u(\cdot)$ is CPA, \eqref{eq:SemUconstraint} implies $Hu(x)\leq h_c$ for all $x\in\T$. Thus, the claim follows from Theorem\;\ref{thm:genOpt}.
\end{proof}

\section{Primary Controller Objectives}\label{sc:PrimaryObjs}
The results in this section are formulated by modifying Instantiation\;\ref{insta:semiProg} to ensure primary goals for the controller. 

\subsection{Stabilization}
The following lemma, improves on \cite[Def\;2,\;Rem\;5]{gieslRevCPA2013} by bounding the convergence rate of $||x(t)||$ above.

\begin{lemma} \label{lem:myExpoStability}
The origin in \eqref{eq:AutSystem}, where $g:\Omega\rightarrow\R^n$ is a Lipschitz map, $\Omega\in\mathfrak{R}^n$, and $g(0)=0$, is locally exponentially stable if there exists a Lipschitz function $V:\Omega\rightarrow\R$ and constants $a,b_1,b_2>0$ satisfying $V(0)=0$, and
\begin{subequations} \label{eq:myExpo}
    \begin{align}
         b_1||x||^a &\leq V(x), \quad \forall x\in\Omega, \textrm{ and} \label{eq:myExpoBound}\\
 D^+V(x) & \leq -b_2 V(x), \quad \forall x\in\Omega\degree \backslash \{0\}. \label{eq:myExpoDecrease}
    \end{align}
\end{subequations}
$V(\cdot)$ is called a Lyapunov function for \eqref{eq:AutSystem}. Further, let $\A=V^{-1}([0,r])\subseteq\Omega$ be in $\mathfrak{R}^n$ for some $r>0$. Then, $||x(t)||\leq\sqrt[^a]{r/b_1}e^{(-b_2/a)(t-t_0)}$, $\forall x(t_0)\in\A\degree$. \qed
\end{lemma}

\begin{proof}
Lyapunov stability is verified by \cite[Def\;2,\;Rem\;5]{gieslRevCPA2013}. Since $\A\in\mathfrak{R}^n$, $x(t_0)\in\A\degree$ implies $V(x(t)) \leq r$ and $x(t)\in\A\degree$ holds for all $t\geq t_0$. Using the Comparison Lemma \cite[Lem 3.4]{khalil}, $V(x(t))\leq V(x(t_0))e^{-b_2(t-t_0)}$ for all $t\geq t_0$. So, $||x(t)||\leq \sqrt[^a]{V(x(0))/b_1} e^{-(b_2/a)(t-t_0)}$, and therefore $||x(t)||\leq\sqrt[^a]{r/b_1}e^{-(b_2/a)(t-t_0)}$.
\end{proof}

The following theorem gives the required modifications to Instantiation\;\ref{insta:semiProg} to find a controller that ensures closed-loop exponential stability of \eqref{eq:controlAffineSystem}, using Lemma\;\ref{lem:myExpoStability}.

\begin{theorem}[Exponential stabilization] \label{thm:cpaStability}
In the Instantiation\;\ref{insta:semiProg}, let $\Omega\in\mathfrak{R}^n$, and f(0) = 0, and augment \eqref{eq:SemiProg} with the constraints $V_0 = u_0 = 0$. If $b_2^\ast>0$, then $V^\ast:\T\rightarrow\R$ is the corresponding CPA Lyapunov function for the closed loop system. Further, let $\A = V^{\ast^{-1}}([0,r])\subseteq\T$ be in $\mathfrak{R}^n$ for some $r>0$. Then $x=0$ is locally exponentially stable for the closed loop system with $||x(t)||\leq\sqrt[^{a^\ast}]{r/b_1^\ast}e^{-(b_2^\ast/a^\ast)(t-t_0)}$ if $x(t_0)\in\A\degree$. \qed
\end{theorem}

\begin{proof}
Using Corollary\;\ref{cor:instantiation}, $b_1^\ast||x||^{a^\ast}\leq D_{i,j}^+V^\ast(x)$ and $D^+V^\ast(x)\leq b_2^\ast V^\ast(x)$ holds for all $x\in\Omega$. Thus, \eqref{eq:myExpoBound} and \eqref{eq:myExpoDecrease} are verified. When $\Omega\in\mathfrak{R}^n$, $b_2^\ast > 0$, $f(0,0)=0$, and $V(0)=u(0)=0$, other conditions of Lemma\;\ref{lem:myExpoStability} are also satisfied, proving the claims.
\end{proof}

Note that $\A$ and $\A\degree$ in Lemma\;\ref{lem:myExpoStability} and Theorem\;\ref{thm:cpaStability} are a positive-invariant set and a \ac{ROA}, respectively since $\A$ is a sublevel set of the Lyapunov function, making the restriction of $V^\ast$ to $\A\degree$ a \ac{CLF}. This is described in the following.

\begin{corollary} \label{cor:myCLF}
If $b^\ast_2>0$ and $\A\in\mathfrak{R}$ are found in Theorem\;\ref{thm:cpaStability}'s optimization, then the restriction of $V^\ast$ to $\A\degree$, that is $V^\ast:\A\degree\rightarrow\R$, is a \ac{CLF} for system \eqref{eq:controlAffineSystem}.
\end{corollary}

\begin{proof}
The claim is verified by the fact that $V^\ast$ is a positive definite function and $u^\ast(x)$ is a feasible point for the following optimization at each $x\in\A\degree$
\begin{equation*}
    \inf_{u\in\U} (D^+V^\ast(x) + b_2^\ast V(x)) \leq 0,
\end{equation*}
as established in Theorem\;\ref{thm:genOpt}'s proof.
\end{proof}

The following corollary parallels Corollary\;\ref{cor:mightFindAPosb2}, giving further conditions for exponential stability in case a non-positive $b_2$ is found.

\begin{corollary} \label{cor:mightFindStabController}
Suppose that $b_2^\ast\leq 0$ is found in Theorem\;\ref{thm:cpaStability}'s optimization, and $\exists\hat{\T}\neq\emptyset$ implying $\hat{b}_2>0$ as defined in Corollary\;\ref{cor:mightFindAPosb2}. If $\A = V^{\ast^{-1}}([0,r])\subseteq\hat{\T}$ is in $\mathfrak{R}^n$ for some $r>0$, then $x=0$ is locally exponentially stable for the closed loop system with $||x(t)||\leq\sqrt[^{a^\ast}]{r/b_1^\ast}e^{-(\hat{b}_2^\ast/a^\ast)(t-t_0)}$ for all $x(t_0)\in\A\degree$. \qed
\end{corollary}

\begin{proof}
Having $\hat{\T}\in\mathfrak{R}^n$ in Corollary\;\ref{cor:mightFindAPosb2} ensures that $0\in\hat{\T}$, making exponential stability possible by Theorem\;\ref{thm:cpaStability}.
\end{proof}

\subsection{Finding Controllable/Stabilizable Sets}
Given the system \eqref{eq:controlAffineSystem} and a target set $\A_1\in\mathfrak{R}^n$ that may or may not be control-invariant, the required modifications to Instantiation\;\ref{insta:semiProg} to search for control-invariant set of feasible states that can reach $\A_1$ are given in this section. The following definitions are needed to establish the results.

\begin{definition}[Controllable/Stabilizable Sets, Ch\,10 {\!\!\cite{BorrelliBook}}] \label{def:controllableSets}
For a given target set $\A_1\subseteq\X$ in \eqref{eq:controlAffineSystem}, a controllable set $\A$ is a control-invariant set, where each state in it can be driven to $\A_1$. If $\A_1$ is control-invariant, the set $\A$ is called a stabilizable set for $\A_1$.
\end{definition}

Note that for the trivial $\A_1=\{0\}$, the set $\A\degree$ in Theorems\;\ref{lem:myExpoStability} and \ref{thm:cpaStability} is a stabilizable set. For non-trivial $\A_1$, finding a controllable/stabilizable set will be established by a modified definition of barrier functions, given next.

\begin{definition} \label{def:myBarrier}
Consider the system \eqref{eq:AutSystem}, where $g:\X\rightarrow\R^n$ is a Lipschitz map. Let a Lipschitz function $V:\Omega\rightarrow\R$, where $\Omega\subseteq\X$ is in $\mathfrak{R}^n$, and the sets $\A,\A_1\in\mathfrak{R}^n$, where $\A\supset\A_1$ is a sublevel set of $V$, and $a,b_1,b_2 > 0$ satisfy
\begin{subequations}
\begin{alignat}{2}
    b_1||x||^a &\leq V(x) \leq V_{\del\A}, \quad && \forall x\in\A, \label{eq:Wc1Closure} \\
    D^+V(x) &\leq -b_2V(x), && \forall x\in (\A\backslash\A_1)\degree, \label{eq:dWc1}
\end{alignat}
\end{subequations}
where $V_{\del\A}$ is the constant value of $V(x)$ on $\del\A$. Then, the restriction of $V(\cdot)$ to $\A\degree$, that is $V:\A\degree\rightarrow\R$, is a barrier function for \eqref{eq:AutSystem}. \qed
\end{definition}

Definition\;\ref{def:myBarrier} modifies the zeroing barrier function definition in \cite{Ames2016} by requiring a positive value for $V$ on the boundary of the set, and allowing $V$ to have positive time derivatives only inside $\A_1$. So, contrary to \cite{Ames2016}, it is not possible for the state to be attracted to $\del\A$. In fact, the following theorem ensures positive-invariance of $\A$ and reachability of $\del\A_1$ from $(\A\backslash\A_1)\degree$. Moreover, it provides a clear upper-bound on the decay rate of the state norm in $(\A\backslash\A_1)\degree$.

\begin{lemma} \label{lem:barrierExistence}
If there exist a function $V(x)$, and sets $\A$, and $\A_1$ satisfying Definition\;\ref{def:myBarrier}, then $\A$ is a positive-invariant set for \eqref{eq:AutSystem}. Further, if $x(t_0)\in (\A\backslash\A_1)\degree$, the state reaches $\del\A_1$ and $||x(t)||\leq \sqrt[^a]{V_{\del\A}/b_1} e^{-(b_2/a)(t-t_0)}$ holds as long as $x(t)$ remains in $(\A\backslash\A_1)\degree$.
\end{lemma}

\begin{proof}
Positive-invariance of $\A$ and reachability of $\del\A_1$ in case $x(t_0)\in(\A\backslash\A_1)\degree$ follow as a special case of \cite[Thm\;2.6]{giesl2015} because $b_2V(x)$ is a positive number and $\A,\A_1\in\mathfrak{R}^n$. Having $x(t_0)\in (\A\backslash\A_1)\degree$ implies $V(x(t_0))\leq V_{\del\A}$ by \eqref{eq:Wc1Closure}. The inequality $||x(t)||\leq \sqrt[^a]{V_{\del\A}/b_1} e^{-(b_2/a)(t-t_0)}$ then follows from \eqref{eq:dWc1} using the Comparison Lemma, \cite[Lem 3.4]{khalil}.
\end{proof}

Note that given a target set $\A_1\in\mathfrak{R}^n$, Lemma\;\ref{lem:barrierExistence} gives sufficient conditions for finding a controllable/stablizable set, $\A\degree\supset\A_1$ for it. The following theorem gives the required modifications to Instantiation\;\ref{insta:semiProg} to synthesize a controller that finds a controllable/stablizable set for a triangulable target set.

\begin{theorem}[Reaching a target] \label{thm:cpaSafety}
Given the two triangulable sets $\A,\A_1\in\mathfrak{R}^n$, where $\A\supset\A_1$, let $\Omega:=\A$ in the Instantiation\;\ref{insta:semiProg}, and suppose that $\del \A_1$ is constrained in $\T$. Denote the set of simplex indices in $\A_1$ by $\I_1=\Set{i\in\IntSet_1^{\mt} | \sigma_i\in\A_1}$ and let $\I_0=\IntSet_1^{m_\T}\backslash\I_1$. Further, let $V_{\del\T}$ be the single unknown value of all $V_x$'s satisfying $x\in\mathbb{E}_{\del\T}$, and $\mathbf{b}=\{b_1,b_2,b_3\}\in\R$. Replace \eqref{eq:SemiVconstraint} by $b_1||x||^a \leq V_x \leq V_{\del\T},\; \forall x\in\Et$, and \eqref{eq:semDv} by $D^+_{i,j}V \leq -b_k V_{x_{i,j}}, \; \forall i\in\I_{k-2}, \; \forall j\in\IntSet_0^n$, where $k\in\{2,3\}$. If $b_2^\ast>0$, then the CPA function $V^\ast:\T\rightarrow\R$ constructed from the elements of $\textbf{V}^\ast$ is a barrier function for the closed loop system, making $\A\degree$ positive-invariant. Further, starting at any $x(t_0)\in(\A\backslash\A_1)\degree$, the state reaches $\del\A_1$ and as long as $x(t)$ remains in $(\A\backslash\A_1)\degree$, $||x(t)||\leq\sqrt[^{a^\ast}]{V_{\del\A}^\ast/b^\ast_1}e^{-(b^\ast_2/a^\ast)(t-t_0)}$ holds. \qed
\end{theorem}

\begin{proof}
To see that replacing \eqref{eq:SemiVconstraint} by $b_1||x||^a \leq V_x \leq V_{\del\T},\; \forall x\in\Et$ does not harm feasibility, note that $V_x=b_1||x||^a$ with any $a,b_1>0$ creates a temporary set $\mathbf{V}^{\textrm{temp}}=\{V_x^{\textrm{temp}}\}_{x\in\Et}$. Then by replacing $V_{\del\T}=\max_{x\in\mathbb{E}_{\del\T}} \mathbf{V}^{\textrm{temp}}$ for all the elements of $\mathbf{V}^{\textrm{temp}}$ corresponding to the vertices on $\del\T$ and keeping all other elements unchanged, $\mathbf{V}$ satisfying $b_1||x||^a \leq V_x \leq V_{\del\T},\; \forall x\in\Et$ is obtained. Using similar arguments to those of Theorem\;\ref{thm:genOpt}'s proof, feasible $\mathbf{L}$, $\mathbf{U}$, $\mathbf{Z}$ can be found. Also, replacing \eqref{eq:semDv} by $D^+_{i,j}V_{x_{i,j}} \leq -b_k V, \; \forall i\in\I_{k-2}, \; \forall j\in\IntSet_0^n$, where $k\in\{2,3\}$, means that $D_{i,j}^+V\leq -b_2V_{x_{i,j}}$ and $D_{i,j}^+V_{x_{i,j}}\leq -b_3$ must hold for all simplexes in $\A\backslash\A_1$ and $\A_1$, respectively. As discussed in Theorem\;\ref{thm:genOpt}'s proof, since $D_{i,j}^+V_{x_{i,j}}$ has finite values for all simplexes, finite $b_2,b_3\in\R$ exist to satisfy these inequalities.

The remainder of the proof is devoted to showing that $V^\ast$ for the closed-loop system verifies Definition\;\ref{def:myBarrier} and Lemma\;\ref{lem:barrierExistence} if $\hat{b}^\ast_2> 0$. Letting $\Omega\coloneqq\A$ in Definition\;\ref{def:myBarrier}, since $V^\ast$ is CPA, $b_1^\ast||x||^{a^\ast} \leq V_x^\ast \leq V_{\del\T},\; \forall x\in\Et$ implies \eqref{eq:Wc1Closure}. By Corollary\;\ref{cor:instantiation},  \eqref{eq:SemiNablaConstraint}--\eqref{eq:semDv} imply \eqref{eq:dWc1}. Also, letting $\Omega\coloneqq(\A\backslash\A_1)\degree$ in Corollary\;\ref{cor:instantiation}, \eqref{eq:dWc1} is verified regardless of $\hat{b}_3$. The upper-bound on the decay rate of the state norm and reachability of $\del\A_1$ follow from Lemma\;\ref{lem:barrierExistence}.
\end{proof}

Even if $b_2^\ast\leq 0$ is found in Theorem\;\ref{thm:cpaSafety}, there might exist a controllable set for the target $\A_1$ on a subset of $\A$ paralleling the idea of Corollary\;\ref{cor:mightFindAPosb2}, as the following.

\begin{corollary} \label{cor:mightFindSafetyController}
Suppose that $b_2^\ast\leq 0$ is found in Theorem\;\ref{thm:cpaSafety}'s optimization, and $\exists\hat{\T}\neq\emptyset$ implying $\hat{b}_2>0$ as defined in Corollary\;\ref{cor:mightFindAPosb2}. If $\hat{\A} = V^{\ast^{-1}}([0,r])\subseteq\hat{\T}$ satisfies $\A_1\subset\hat{\A}\subset\A$ for some $r>0$, then $\hat{\A}$ is positive-invariant, and starting at any $x(t_0)\in(\hat{\A}\backslash\A_1)\degree$, as long as $x(t)$ remains in $(\hat{\A}\backslash\A_1)\degree$, $||x(t)||\leq\sqrt[^{a^\ast}]{V_{\del\hat{\A}}^\ast/b^\ast_1}e^{-(\hat{b}^\ast_2/a^\ast)(t-t_0)}$ holds. \qed
\end{corollary}

\section{Enlarging the Region of Attraction}\label{sc:EnlargingROA}
As established by Theorem\;\ref{thm:cpaStability}\;(Exponential stabilization), the set $\A$, a sublevel set of the obtained Lyapunov function provides a \ac{ROA} for the closed-loop system. However, it might be quite small relative to the triangulated set on which the corresponding optimization is solved because $\A$ must be entirely inside the triangulation. When $b_2^\ast>0$ on a triangulation is found, it indicates that larger \ac{ROA} might exist even though the largest sublevel set insided the triangulation is small. On the other hand, the optimization in Theorem\;\ref{thm:cpaSafety}\;(Reaching a target) is forced to make the boundary of the triangulation a positive-invariant set. This idea can be used to find a larger \ac{ROA} when the target set is the origin. Further, once a \ac{ROA} is found, it can be used as the control-invariant target set to find a large stabilizable set that contains it by Theorem\;\ref{thm:cpaSafety}\;(Reaching a target). These two ideas are explored in this section by calling them single-stage and multi-stage design. The latter not only enlarges the \ac{ROA}, but also has useful practical properties.

\subsection{Single-Stage Design}
By assuming $\A_1=\emptyset$ and $V_0=0$ in Theorem\;\ref{thm:cpaSafety}\;(Reaching a target), the following theorem tries to make the boundary of the triangulation a \ac{ROA}. Since it might not be possible to do so, a corollary that parallels Corollary\;\ref{cor:mightFindStabController} will be used to find the largest \ac{ROA} it can find inside the triangulation.

\begin{theorem}[Single-stage ROA] \label{thm:singleStage}
Given the polytope $\A\in\mathfrak{R}^n$, let $\Omega:=\A$ in the Instantiation\;\ref{insta:semiProg}. Let $V_{\del\T}$ be the single unknown value of all $V_x$'s satisfying $x\in\mathbb{E}_{\del\T}$. Replace \eqref{eq:SemiVconstraint} by $b_1||x||^a \leq V_x \leq V_{\del\T},\; \forall x\in\Et$, and \eqref{eq:semDv}. If $b_2^\ast>0$, then $V^\ast:\T\rightarrow\R$ is the corresponding CPA Lyapunov function for the closed loop system, making $x=0$ locally exponentially stable for the closed loop system, where  $||x(t)||\leq\sqrt[^{a^\ast}]{V_{\del\A}^\ast/b^\ast_1}e^{-(b^\ast_2/a^\ast)(t-t_0)}$ holds for all $x(t_0)\in\A\degree$. \qed
\end{theorem}

\begin{proof}
The claims follow from Theorems\;\ref{thm:cpaStability} and \ref{thm:cpaSafety}.
\end{proof}

In case a non-positive $b_2^\ast$ is found, the following corollary might find a nonempty set $\A$, paralleling Corollary\;\ref{cor:mightFindAPosb2}. Although it has the same logic of Corollary\;\ref{cor:mightFindStabController}, for completeness, it is stated for completeness.  

\begin{corollary} \label{cor:mightFindSingleStage}
Suppose that $b_2^\ast\leq 0$ is found in Theorem\;\ref{thm:singleStage}'s optimization, and $\exists\hat{\T}\neq\emptyset$ implying $\hat{b}_2>0$ as defined in Corollary\;\ref{cor:mightFindAPosb2}. If $\A = V^{\ast^{-1}}([0,r])\subseteq\hat{\T}$ is in $\mathfrak{R}^n$ for some $r>0$, then $x=0$ is locally exponentially stable for the closed loop system with $||x(t)||\leq\sqrt[^{a^\ast}]{r/b_1^\ast}e^{-(\hat{b}_2^\ast/a^\ast)(t-t_0)}$ for all $x(t_0)\in\A\degree$. \qed
\end{corollary}

As will be shown in Section\;\ref{sc:Examples}, the proposed single-stage design might be used to blindly search for a large \ac{ROA}, but it usually can do so on a very refined triangulation. Thus, expanding the \ac{ROA} via a multi-stage design is proposed next.

\subsection{Multi-Stage Design}
When $\A_1\neq\emptyset$ in Theorem\;\ref{thm:cpaSafety}\;(Reaching a target), a large positive-invariant set can be found because its optimization ignores the time derivative of $V(\cdot)$ inside $\A_1$. Moreover, one can easily verify that $V(\cdot)$ inside $\A_1$ can be trivially selected by connecting each vertex on $\A_1$ to the origin and substituting its value in the appropriate constraints. This independence from what happens inside $\A_1$ means that reaching $\del\A_1$ from the points in $(\A\backslash\A_1)\degree$ can be still ensured even if \eqref{eq:SemiProg} is solved on a hollowed triangulated region that has $\del\A_1$ as its inner constraint surface. If $\A_1\degree$ is itself a \ac{ROA} found by \textit{ex} Theorem\;\ref{thm:cpaStability}\;(Exponential stabilization) or \ref{thm:singleStage}\;(Single-stage ROA), finding $\A\supset\A_1$ means that a larger \ac{ROA} is obtained. This motivates a multi-stage design where in each stage, a hollowed region surrounds last-stage's positive-invariant set. In the proposed multi-stage design, the \ac{CPA} choice for Lyapunov-like functions and controllers makes it possible to modify optimizations that search for them to either stitch them to their corresponding functions found previously in the inner region, or let each/both of them be piecewise \ac{CPA} functions if discontinuous functions along the boundary of the inner region are preferred.

There are two main advantages for a multi-stage design. Note that as the number of simplexes in a triangulation increases, so does the number of constraints in Instantiation\;\ref{insta:semiProg}, complicating the optimizations in Theorems\;\ref{thm:cpaStability} and \ref{thm:singleStage} more complicated. However, each stage of the multi-stage design may have far fewer simplexes, alleviating taxing computations. Moreover, each stage of the multi-stage design can ensure its own secondary objective. This can be, for instance imposing different convergence rates at each stage.

The following theorem describes each design stage. The variables related to stage $k$ are denoted by superscript $[k]$.

\begin{theorem}[Multi-stage ROA] \label{thm:multiStage}
Let $\mathbf{y}^{[k]} =  [\mathbf{V}^{[k]}, \mathbf{L}^{[k]},$ $\mathbf{U}^{[k]}, \mathbf{Z}^{[k]}, a^{[k]}, \mathbf{b}^{[k]}]$, where $k\in\mathbb{N}$, satisfy Theorem\;\ref{thm:cpaStability}\;(Exponential stabilization) or \ref{thm:singleStage}\;(Single-stage ROA) on $\T^{[k]}$, making the origin exponentially stable in the interior of $\A^{[k]}\in\mathfrak{R}^n$. Let $\T\A^{[k]} = \T^{[k]} \cap \A^{[k]}$, and  $\Omega_{\textrm{temp}}\supset\A^{[k]}$ be a triangulable set in $\mathfrak{R}^n$. In the Instantiation\;\ref{insta:semiProg}, let $\Omega \coloneqq \A$, where $\A =\overline{\Omega_{\textrm{temp}}\backslash\A^{[k]}}$, and denote its triangulation by $\T$. Denote the set of vertices on the outer and inner  boundaries of $\Omega$ by $\mathbb{E}_{\del\Omega^{\textrm{out}}}$ and $\mathbb{E}_{\del\A^{[k]}}$, respectively. Replace \eqref{eq:SemiVconstraint} by $b_1||x||^a \leq V_x \leq V_{\del\Omega^{\textrm{out}}},\; \forall x\in\mathbb{E}_{\del\Omega^{\textrm{out}}}$, where $V_{\del\Omega^{\textrm{out}}}$ is the single unknown value of $V(x)$ on $\del\Omega^{\textrm{out}}$. Moreover, augment \eqref{eq:SemiProg} with the constraints $V_x = V_x^{[k]}, \; \forall x\in\mathbb{E}_{\del\A^{[k]}}$ and $u_x = u_x^{[k]}, \; \forall x\in\mathbb{E}_{\del\A^{[k]}}$. If $b_2^\ast > 0$, 
then $\mathbf{y}^{[k+1]}$ satisfies Theorem\;\ref{thm:singleStage} on $\T^{[k+1]}\coloneqq\T\A^{[k]}\cup\T$, making the origin locally exponentially stable in the interior of $\A^{[k+1]} \coloneqq \A^{[k]} \cup \A $, where $\mathbf{V}^{[k+1]} \coloneqq \{V_x^\ast\}_{x\in\mathbb{E}_{\T}} \cup  \{V_x^{[k]}\}_{x\in\mathbb{E}_{\T\A^{[k]}}}$, and $\mathbf{L}^{[k+1]}\coloneqq\{l_i^\ast\}_{i=1}^{m_{\T}}\cup \{l_i^{[k]}\}_{i=1}^{m_{\T\A^{[k]}}}$, and $\mathbf{U}^{[k+1]}\coloneqq\{u_x^\ast\}_{x\in\mathbb{E}_{\T}} \cup  \{u_x\}_{x\in\mathbb{E}_{\T\A^{[k]}}}$, and $a^{[k+1]} \coloneqq \min\{a^\ast,a^{[k]}\}$, and $\mathbf{b}^{[k+1]} \coloneqq \{\min\{b_1^\ast,b_1^{[k]}\},\min\{b_2^\ast,b_2^{[k]}\}\}$. Moreover, $||x(t)||\leq\sqrt[^{a^{[k+1]}}]{V_{\del\Omega^{\textrm{out}}}/b_1^{[k+1]}}e^{-(b_2^{[k+1]}/a^{[k+1]})(t-t_0)}$ if $x(t_0)\in\A\degree$. \qed
\end{theorem}

\begin{proof}
Using the definition of $\mathbf{y}^{[k+1]}$, the proof follows from that of Theorem\;\ref{thm:singleStage} since $V_{\del\Omega^{\textrm{out}}}$ is a level set of the \ac{CPA} Lyapunov function constructed from the elements of $\mathbf{V}^{[k+1]}$. The given upper-bound on the decay rate of the state norm in $\A\degree$ follows from Theorem\;\ref{thm:cpaSafety}.
\end{proof}

\begin{figure}
    \centering
    \includegraphics[width=8.65cm]{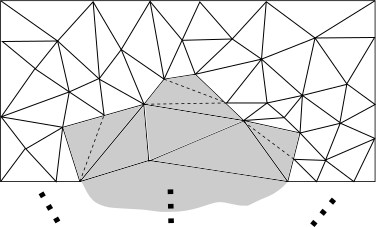}
    \caption{The sets $\A^{[k]}$ (gray), $\Omega$ (white), and their triangulations. The mismatch between vertices when combined makes no ambiguity in evaluating $V(\cdot)$, $u(\cdot)$, and $D^+V(\cdot)$ since $V(\cdot), u(\cdot)$ are CPA. The simplexes with dashes make the combination a typical triangulation without adding any new information.}
    \label{fig:nonRegular}
\end{figure}

As the sets $\A^{[k]}$ and $\T$ are put together in stage $k+1$ in Theorem\;\ref{thm:multiStage}\;(Multi-stage ROA), the result may not look like a triangulation, because the vertices on the inner boundary of $\Omega$ and $\del\A^{[k]}$ may not match. However, since these boundaries are same, and both $V(\cdot), u(\cdot)$ are CPA functions, there is no ambiguity in determining their values on $\del\A^{[k]}$. In fact, the apparent mismatch between $\T$ and the triangulation of $\A^{[k]}$ can be easily resolved by connecting appropriate vertices of the two to generate some redundant simplexes that carry no new information as depicted in Fig.\;\ref{fig:nonRegular}.

\begin{remark} \label{rem:discontVu}
Theorem\;\ref{thm:multiStage}\;(Multi-stage ROA) ensures that the CPA functions of consecutive stages remain continuous since they are stitched together. Relaxations of Theorem\;\ref{thm:multiStage} can be expressed by allowing CPA functions of the Lyapunov-like functions or controllers of consecutive stages to be discontinuous along the boundary of $\A^{[k]}$'s, making them piecewise CPA. This is easily done by eliminating either or both of the constraints $V_x=V_x^{[k]}$, $\forall x\in\mathbb{E}_{\del\A^{[k]}}$ and $u_x=u_x^{[k]}$, $\forall x\in\mathbb{E}_{\del\A^{[k]}}$. Since each $\A^{[k]}$ contains $\A^{[k-1]}$, having finite number of stages means finite number of switches in the Lyapunov function (or in the controller), ensuring overall stability.
\end{remark}

\begin{corollary} \label{cor:mightFindmultiController}
Suppose that $b_2^\ast\leq 0$ is found in Theorem\;\ref{thm:multiStage}'s optimization, and $\exists\hat{\T}\neq\emptyset$ implying $\hat{b}_2>0$ as defined in Corollary\;\ref{cor:mightFindAPosb2}. If $\hat{\A}^{[k+1]} = V^{[k+1]^{-1}}([0,r])\subseteq\hat{\T}$ satisfies $\A^{[k]}\subset\hat{\A}^{[k+1]}\subset\Omega_{\textrm{temp}}$ and $\hat{\A}^{[k+1]}\in\mathfrak{R}^n$ for some $r>0$, then $x=0$ is locally exponentially stable for the closed loop system with $||x(t)||\leq\sqrt[^{a^{[k+1]}}]{V_{\del\hat{A}^{[k+1]}}/b_1^{[k+1]}}e^{-(b_2^{[k+1]}/a^{[k+1]})(t-t_0)}$, if $x(t_0)\in\hat{\A}^{[k+1]}\degree$, where $V_{\del\hat{A}^{[k+1]}}$ is the value of $V^{[k+1]}(\cdot)$ on $\del\A^{[k+1]}$, and $b_2^{[k+1]} \coloneqq \min\{\hat{b}_2^\ast,b_2^{[k]}\}$.  \qed
\end{corollary}

\begin{proof}
Follows similarly to those of Corollary\;\ref{cor:mightFindStabController} and \ref{cor:mightFindSafetyController}.
\end{proof}

For the first stage in the proposed multi-stage design, three options are as follows. One, using Theorem\;\ref{thm:cpaStability}\;(Exponential stabilization). Second, using Theorem\;\ref{thm:singleStage}\;(Single-stage ROA). Third, if the state transition matrix of the linearized system around the origin is Hurwitz, one can design a linear controller for it. Then, using \cite[Thm\;5]{gieslRevCPA2013}, find a sublevel set $\A$ for the closed-loop system's Lyapunov function, making sure that the state and input constrains are satisfied in it.

 \section{Iterative Algorithm}\label{sc:IterativeAlgos}
As discussed in Theorems\;\ref{thm:cpaStability}--\ref{thm:multiStage}, formulating Lyapunov-like functions depends on finding  $b_2^\ast>0$, making it the priority over all other objectives. Choosing a cost function that weighs increasing $b_2$ against other objectives is a bad choice because no useful controller is formulated until $b_2>0$ is found. Once it is found, it can be kept fixed during further optimizations to achieve other performance objectives. However, the proposed optimizations in Theorems\;\ref{thm:cpaStability}--\ref{thm:multiStage} have nonlinear constraints that increase in number alongside the simplexes. This section gives an iterative algorithm that implements the discussed strategy using iterative \ac{SDP}s, convexifying the optimizations with conservatism.

Since the optimizations in Theorems\;\ref{thm:cpaStability}--\ref{thm:multiStage} were formulated using Instantiation\;\ref{insta:semiProg}, this section gives an algorithm that iteratively searches for $b_2>0$ in it using a sequence of \ac{SDP}s. If a sufficiently large $b_2>0$ is found, the algorithm fixes it, and then optimizes other performance objectives. The following theorem formulates each iteration. The required modifications to formulate similar \ac{SDP}s for the optimizations in Theorems\;\ref{thm:cpaStability}--\ref{thm:multiStage} are inferred from their corresponding changes to Instantiation\;\ref{insta:semiProg}, but their initilizations will be discussed separately.

\begin{theorem}[Iterative SDP] \label{thm:semiProg2}
Consider \eqref{eq:SemiProg}, where $a\geq1$ is a fixed number. Let $\underline{\mathbf{y}}=[\underline{\mathbf{V}}, \underline{\mathbf{L}}, \underline{\mathbf{U}}, \underline{\mathbf{Z}}, a,  \underline{\mathbf{b}}]$ satisfy \eqref{eq:SemiVconstraint}--\eqref{eq:semDv}. Consider the following optimization.
\begin{subequations} \label{eq:SemiProg2} 
    \begin{alignat}{2}
        &\delta{\mathbf{y}^\ast} = \argmin_{\delta\mathbf{y} =  [\delta\mathbf{V}, \delta\mathbf{L}, \delta\mathbf{U}, \delta\mathbf{Z}, 0, \delta\mathbf{b}]} && J(\underline{\mathbf{y}}+\delta\mathbf{y}) \nonumber \\
        &\textrm{s.t.} && \nonumber \\
        & \underline{b}_1 + \delta b_1 > 0, && \label{eq:Semi2V1b0Constraint} \\
        & (\underline{b}_1+\delta b_1)||x||^a \leq \underline{V}_x +  \delta V_x, && \forall x\in\Et, \label{eq:SemiVconstraint2} \\
        & |{\nabla\underline{V}}_i + \delta {\nabla V}_i| \leq \underline{l}_i + \delta l_i, \quad && \forall i\in\IntSet_1^{\mt},  \label{eq:SemiNablaConstraint2} \\
        & H (\underline{u}_x + \delta u_x) \leq h_c, && \forall x \in \Et, \label{eq:SemUconstraint2} \\
        &|\nabla{\underline{u}^{(s)}}_i {+} \delta \nabla{u^{(s)}}_i| \leq \underline{z}_i {+} \delta z_i, \;\; && \forall i\in\IntSet_1^{\mt}, \; \forall s\in\IntSet_1^m, \label{eq:SemDuConstraint2} \\
        & P_{i,j} \leq 0, && \forall i\in\IntSet_1^{\mt}, \; j{=}0, \label{eq:2SemDv1} \\
        & Q_{i,j} \leq 0, && \forall i\in\IntSet_1^{\mt}, \; \forall j\in\IntSet_1^n, \label{eq:2SemDv2}
        \end{alignat}
\end{subequations}

\noindent where $\delta{\nabla V}_i{=}X_i^{-1}\delta\bar{V}_i$, $\delta{\nabla u^{(s)}}_i{=}X_i^{-1}\delta\bar{u}_i$ as in Remark\;\ref{rem:nablaLinear},
\begin{equation}
P_{i,j} = \begin{bmatrix} \hat{\phi}_{i,j} & \ast & \ast & \ast & \ast \\
                          \delta{\nabla{V}}_i & -2I_n & \ast & \ast & \ast \\
                          G(x_{i,j})\delta u_{x_{i,j}} & 0 & -2I_n & \ast & \ast \\
                          \delta V_{x_{i,j}} & 0 & 0 & -2 & \ast \\
                          \delta b_2 & 0 & 0 & 0 & -2 \end{bmatrix},
\end{equation}

\begin{align}
    \hat{\phi}_{i,j} = & ({\nabla \underline{V}}_i + \delta{\nabla V}_i)^\intercal(f(x_{i,j}) + G(x_{i,j}) \underline{u}_{x_{i,j}}) + \ldots \nonumber \\
    & {\nabla \underline{V}}_i^\intercal G(x_{i,j})\delta u_{x_{i,j}} + \mu_ic_{i,j}1_n^\intercal (\underline{l}_i+\delta l_i)  + \ldots \nonumber \\
    & \eta_ic_{i,j}\left(( \underline{z}_i+\delta z_i)1_n^\intercal  \underline{l}_i + \underline{z}_i 1_n^\intercal\delta l_i \right) + \ldots  \nonumber \\
    & b_2 ( \underline{V}_{x_{i,j}}  + \delta V_{x_{i,j}}) +  \underline{V}_{x_{i,j}}\delta b_2, \textrm{ and} \label{eq:phiHat} 
\end{align}

\begin{equation}
Q_{i,j} = \begin{bmatrix} P_{i,j} & \ast & \ast\\
                          1_n^\intercal\delta l_i & \frac{-2}{\eta_i c_{i,j}} & \ast \\
                          \delta z_i & 0 & \frac{-2}{\eta_i c_{i,j}} \end{bmatrix},
\end{equation}

\noindent and $c_{i,j}$ is given in \eqref{eq:betaAndc}, and $\mu_i, \eta_i$ are given in \eqref{eq:affineB}. Then, $\underline{\mathbf{y}}+\delta\mathbf{y}^\ast$ is a feasible point for \eqref{eq:SemiProg}, and $J(\underline{\mathbf{y}}+\delta\mathbf{y}^\ast)\leq J(\underline{\mathbf{y}})$. \qed
\end{theorem}

\begin{proof}
To see that \eqref{eq:SemiProg2} is feasible, observe that $\delta\mathbf{y}{=}0$ satisfies \eqref{eq:SemiProg2} since in this case, \eqref{eq:SemiProg2} is equivalent to \eqref{eq:SemiProg} with $\mathbf{y}:=\underline{\mathbf{y}}$. In fact, \eqref{eq:2SemDv1}--\eqref{eq:2SemDv2} are the convexified versions of \eqref{eq:semDv}. To see this, recall that $w^\intercal v {\leq} 1/2(w^\intercal w {+}v^\intercal v)$ for any vectors $v,w$ with the same dimension. Applying this fact with $(v,w){=}(\delta \nabla{V}_i, G(x_{i,j}) \delta u_{x_{i,j}})$, $(v,w){=}( \delta z_i, 1_n^\intercal\delta l_i)$, and $(v,w){=}(\delta V_{x_{i,j}}, \delta b_2)$ shows that by the Schur complement, \eqref{eq:2SemDv1} is implied when $j{=}0$, since $c_{i,j}$ is zero in this case, and \eqref{eq:2SemDv2} is implied when $j{\neq} 0$. Finally, $J(\underline{\mathbf{y}} {+} \delta \mathbf{y}) \leq J(\mathbf{y})$ because otherwise $\delta\mathbf{y}=0$ would be a better, feasible solution.
\end{proof}

\begin{remark}
In case $G(x)$ in \eqref{eq:controlAffineSystem} is a constant matrix, \eqref{eq:2SemDv1} must be used for all $j\in\IntSet_0^n$ because in this case, $\eta_i=0$ in \eqref{eq:affineB}. Also, \eqref{eq:SemDuConstraint} and \eqref{eq:SemDuConstraint2} are not needed. \qed
\end{remark}

Starting with a feasible point of \eqref{eq:SemiProg},  Theorem\;\ref{thm:semiProg2}\;(Iterative SDP) can be used repeatedly to potentially decrease the values of the cost function. Note that by  letting $\underline{b}_1+\delta b_1$ be greater than or equal to a small positive number, and a linear cost function, \eqref{eq:SemiProg2} is a SDP in the standard format. The small positive number must be kept constant in the later iterations. Two feasible initialization points for \eqref{eq:SemiProg} are given next.

\begin{initialization} \label{initialization:random}
Choosing $a\geq1$ and $b_1>0$, let $V_x=b_1||x||^a$, $\forall x{\in}\Et$. Assign admissible $u_x$ for all $x{\in}\Et$. They can be random. Compute $l_i{=}|\nabla{V}_i|$ and $z_i{=}|\nabla{u^{(s)}}_i|$ for all $i\in\IntSet_1^{\mt}$ as in Remark\;\ref{rem:nablaLinear}. Finally, find the largest $b_2$ satisfying \eqref{eq:semDv} in all simplexes. \qed
\end{initialization}

\begin{initialization} \label{initialization:LQR}
Linearize \eqref{eq:controlAffineSystem} around the origin. Design a LQR controller, and find the corresponding quadratic Lyapunov function, $x^\intercal\hat{P}x$. Sample $x^\intercal\hat{P}x$ at the vertices of $\T$ to find $\mathbf{V}$, and let $a{=}2$ and $b_1$ be equal to the smallest eigenvalue of $\hat{P}$. Sample the LQR controller at the vertices of $\T$ to form $\mathbf{U}^{\textrm{LQR}}=\{u_x^{\textrm{LQR}}\}_{x\in\Et}$. Divide each element of $\mathbf{U}^{\textrm{LQR}}$ by an appropriate positive number so that the result, $\mathbf{U}=\{u_x\}_{x\in\Et}$, has admissible values for all vertices. Compute $l_i{=}|\nabla{V}_i|$ and $z_i{=}|\nabla{u^{(s)}}_i|$ for all $i{\in}\IntSet_1^{\mt}$ as in Remark\;\ref{rem:nablaLinear} using the computed values of $V_x$ and $u_x$, respectively. Finally, find the largest $b_2$ satisfying \eqref{eq:semDv} in all simplexes. \qed
\end{initialization}

Given a triangulation and a convex cost function $\hat{J}(\textbf{V},\textbf{L},\textbf{U},\textbf{Z},b_1)$, and a feasible initial point $\underline{\mathbf{y}}$, the procedure for finding a positive $b_2^\ast$ and minimizing $J(\cdot)$ in Instantiation\;\ref{insta:semiProg} is given in Algorithm\;\ref{alg:cpaControl}. It iteratively increases $b_2$ until it is positive. Since $e^{-(\sfrac{b_2}{a})t}$ is proportional to the state norm's upper-bound when $a{>}0$ is fixed, increasing $b_2{>}0$ can continue until a desired decay rate is ensured. Then, by fixing $b_2$'s value, $\hat{J}(\cdot)$ is iteratively minimized. Both of the loops can be terminated in lines \ref{line:term1} and \ref{line:term2} if a predefined maximum number of iterations is reached. If a sufficiently large positive $b_2$ cannot be found, triangulation refinement, discussed later, is needed. 

\begin{algorithm}[h!]
	\caption{CPA control design on a fixed triangulation}
    \label{alg:cpaControl}
	\begin{algorithmic}[1]
	   \Require Instantiation\;\ref{insta:semiProg}, and a convex $\hat{J}(\hat{\mathbf{y}})$, where $\hat{\mathbf{y}}=[\textbf{V},\textbf{L},\textbf{U},\textbf{Z},b_1]$
	   \Ensure Sufficiently large $b_2>0$ and $\hat{\mathbf{y}}$
	   \State $\underline{\mathbf{y}} \coloneqq $ a feasible point of \eqref{eq:SemiProg} (using Initialization\;\ref{initialization:random} or \ref{initialization:LQR}) \label{line:initialization}
	   \State $J \coloneqq -b_2$ \Comment{since $b_2$ is to be maximized}
	   \Repeat{}
	       \State Use Theorem\;\ref{thm:semiProg2}\;(Iterative SDP) and update $\hat{\mathbf{y}}$ \label{line:usingTheorem1st}
	   \Until{$b_2 > 0$ is large enough OR $b_2$ is not changing} \label{line:term1}
	   \If{$b_2>0$ is found} \label{line:continueStart}
	       \State Fix $b_2$, and let $J\coloneqq\hat{J}(\cdot)$
	       \Repeat{}
	           \State Use Theorem\;\ref{thm:semiProg2}\;(Iterative SDP) and update $\hat{\mathbf{y}}$ \label{line:usingTheorem2nd}
	       \Until{$J$ is sufficiently small OR $J$ is not changing} \label{line:term2}
	       \State Return $b_2$ and $\hat{\mathbf{y}}$
	   \EndIf \label{line:continueEnd}
	\end{algorithmic}
\end{algorithm}


Once Algorithm\;\ref{alg:cpaControl} is terminated at line\;\ref{line:term2}, or \ref{line:term1} if no other improvement is needed, the returned $\hat{\mathbf{y}}$ can serve as an initial guess for the non-convex optimization \eqref{eq:SemiProg} with a nonlinear cost function $J(\cdot)$, as a final attempt to boost the performance. 

By appropriate modifications that account for differences in the constraints of Instantiation\;\ref{insta:semiProg}'s optimization and those of Theorems'\;\ref{thm:cpaStability}--\ref{thm:multiStage}, iterative versions of Theorem\;\ref{thm:cpaStability}\;(Regulatoin), Theorem\;\ref{thm:cpaSafety}\;(Reaching a target), Theorem\;\ref{thm:singleStage}\;(Single-stage ROA), and Theorem\;\ref{thm:multiStage}\;(Multi-stage ROA) are obtained. Then, provided that they are correctly initialized, Algorithm\;\ref{alg:cpaControl} can be used to find a large enough $b_2>0$ and achieve further objectives for any of them, where one of the corresponding Corollaries\;\ref{cor:mightFindStabController}, \ref{cor:mightFindSafetyController}, \ref{cor:mightFindSingleStage}, or \ref{cor:mightFindmultiController} can be used in line\;\ref{line:usingTheorem1st} to see if a positive $\hat{b}_2$ exists each time a non-positive $b_2$ is found. Once desired objectives are met, the positive-invariant set $\A$ can be also found. Modifying Initializations\;\ref{initialization:random} and \ref{initialization:LQR}, the next section gives feasible points for Theorems\;\ref{thm:cpaStability}--\ref{thm:multiStage} to be used in line\;\ref{line:initialization} of Algorithm\;\ref{alg:cpaControl} when designing by any of them.

\subsection{Specialized Initializations}
Here, for each of Theorems\;\ref{thm:cpaStability}--\ref{thm:multiStage}, two feasible points are given, allowing their iterative versions to be used in Algorithm\;\ref{alg:cpaControl} in lines\;\ref{line:usingTheorem1st} and \;\ref{line:usingTheorem2nd} once properly initialized in line\;\ref{line:initialization}.

\subsubsection{For Theorem\;\ref{thm:cpaStability}\;(Exponential stabilization)'s Optimization}
\begin{initialization}
Use Initialization\;\ref{initialization:random}, and let $u_0=0$. \qed
\end{initialization} 

\begin{initialization}
Same as Initialization\;\ref{initialization:LQR}. \qed
\end{initialization}

\subsubsection{For Theorem\;\ref{thm:cpaSafety}\;(Reaching a target)'s Optimization}
\begin{initialization} \label{init:safety1}
Choosing $a\geq1$ and $b_1>0$, let $V_x=b_1||x||^a$ for $\forall x\in\Et\backslash\mathbb{E}_{\del\T}$, and $V_{\del\T} = \max_{x\in\Et} b_1||x||^a$. The rest is identical to Initialization\;\ref{initialization:random} \qed
\end{initialization}

\begin{initialization} \label{init:safety2}
Use Initialization\;\ref{initialization:LQR} to find $\mathbf{V}=\{V_x\}_{x\in\Et}$. Replace the elements of $\mathbf{V}$ corresponding to boundary vertices with $V_{\del\T}=\max_{x\in\Et}V_x$. The rest is identical to Initialization\;\ref{initialization:LQR}. \qed
\end{initialization}

\subsubsection{For Theorem\;\ref{thm:singleStage}\;(Single-stage ROA)'s Optimization}
\begin{initialization}
Use Initialization\;\ref{init:safety1}, and let $u_0=0$. \qed
\end{initialization}

\begin{initialization}
Same as Initialization\;\ref{init:safety2}. \qed
\end{initialization}

\subsubsection{For Theorem\;\ref{thm:multiStage}\;(Multi-stage ROA)'s Optimization}
\begin{initialization}
Let $a \coloneqq a^{[k]}$, and $b_1 \coloneqq \min_{ x\in\mathbb{E}_{ \del\A^{[k]} }  }  V_{\del\A^{[k]}} /  ||x||^a$. Assign $V_{\del\A^{[k]}}$ to all elements of $\{V_x\}_{x\in\mathbb{E}_{\del\A^{[k]}}}$, and let $V_x = b_1||x||^a$ for all $x\in\mathbb{E}_{\T\backslash\del\A^{[k]}}$ to obtain $\mathbf{V}^{\textrm{temp}}=\{V_x\}_{x\in\Et}$. Now replace all elements in $\mathbf{V}^{\textrm{temp}}$ corresponding to vertices in $\mathbb{E}_{\del\Omega_{\textrm{out}}}$ with $V_{\del\Omega^{\textrm{out}}} \coloneqq \max_{x\in\Et} V_x$. The result is $\{V_x\}_{x\in\Et}$. The rest is identical to Initialization\;\ref{initialization:random}. \qed
\end{initialization}

\begin{initialization}
Linearize \eqref{eq:controlAffineSystem} around the origin. Design a LQR controller, and find the corresponding quadratic Lyapunov function, $x^\intercal\hat{P}x$. Sample $x^\intercal\hat{P}x$ at the vertices of $\T$ to find $\mathbf{V}^{\textrm{temp}}$, and let $a{=}2$ and $b_1$ be equal to the smallest eigenvalue of $\hat{P}$. Replace all elements of $\mathbf{V}^{\textrm{temp}}$ corresponding to vertices in $\del\Omega^{\textrm{out}}$ and $\del\A^{[k]}$ by $V_{\del\Omega^{\textrm{out}}} = \max_{x\in\Et} V_x$ and $V_{\del\A^{[k]}} = \min_{x\in\Et}  V_x$, respectively to obtain $\mathbf{V}=\{V_x\}_{x\in\Et}$. The rest is identical to Initialization\;\ref{initialization:LQR} except for after finding $\mathbf{U}$, where its elements that correspond to vertices in $\del\A^{[k]}$ should be replaced by corresponding values from the \ac{CPA} controller $u^{[k]}(\cdot)$. \qed
\end{initialization}

\section{Minimum-Norm Controllers}\label{sc:MinNormControl}
Using Algorithm\;\ref{alg:cpaControl}, a stabilizing controller that ensures a desired upper-bound for the the decay rate of the state norm can be found at line\;\ref{line:term1} using iterative versions of any of Theorems\;\ref{thm:cpaStability}, \ref{thm:singleStage}, or \ref{thm:multiStage}. Although such a controller is guaranteed to be feasible at all times when the system is initialized in the corresponding positive-invariant set, its norm can be further reduced without losing stability or safety. This section suggests two notions of achieving the minimum-norm for the controller. The first one continues the offline design in Algorithm\;\ref{alg:cpaControl} by minimizing a quadratic cost function. The second one takes the \ac{CLF} obtained at line\;\ref{line:term1} or \ref{line:term2} of Algorithm\;\ref{alg:cpaControl}, and seeks the point-wise minimum-norm controller using online \ac{QP}. Having minimum-norm property at all points and online computations poses a trade-off between the two proposed minimum-norm controllers. Moreover, while Lipschitz property for the offline designed controller is guaranteed, it is not the case for the \ac{QP}-based controller \cite{Ames2016}.

\subsection{Continuing Offline Design}
By Definition\;\ref{def:CPAfunction}, the CPA controller is fully defined by its value at the vertices of a triangulation. After a stabilizing controller results from $b_2>0$ in any of Theorems\;\ref{thm:cpaStability}--\ref{thm:multiStage}, finding a minimum-norm controller can be a secondary objective of further offline optimizations. This can be done by fixing the obtained $a$ and $b_2$, and minimizing the quadratic cost function, $\hat{J}=\mathbf{U}^\intercal\mathbf{U}$. Since the $\mathbf{U}$ contains the values of the \ac{CPA} $u(\cdot)$ at all vertices, minimizing $\mathbf{U}^\intercal\mathbf{U}$ results a controller that has minimum-norm property at all the vertices. Since $u(\cdot)$ is \ac{CPA}, at each point in the interior of any simplex, $u(x)$ will be a linear interpolation of minimum-norm values at the vertices of that simplex. Minimizing $\mathbf{U}^\intercal\mathbf{U}$ can be done as a non-convex optimization or by iterative \ac{SDP}s as follows. It can be considered as one candidate for lines\;\ref{line:continueStart}--\ref{line:continueEnd} of Algorithm\;\ref{alg:cpaControl}.

Let $\underline{\mathbf{y}}=[\underline{\mathbf{V}}, \underline{\mathbf{L}}, \underline{\mathbf{U}}, \underline{\mathbf{Z}}, \underline{a},\underline{\mathbf{b}}]$ be a feasible point of \eqref{eq:SemiProg}, where $\underline{b}_2>0$ is found by any of Theorems\;\ref{thm:cpaStability}--\ref{thm:multiStage}. Let  $\tilde{\mathbf{y}}=[\mathbf{V},\mathbf{L},\mathbf{U},\mathbf{Z},b_1,\phi]$, where $\phi\in\R$ be the unknowns on the same triangulation in which $\underline{\mathbf{y}}$ was found. Fixing $\underline{b}_2$ and $\underline{a}$, the search for the controller that is minimum-norm in all vertices can be formulated as the following \ac{SDP}.

\begin{subequations} \label{eq:minNormSDP} 
    \begin{alignat}{2}
        \tilde{\mathbf{y}}^\ast &= \argmin_{\tilde{\mathbf{y}}} \;\; \phi  \nonumber \\
        \textrm{s.t.} \;\; & \phi \geq 0, \label{eq:minNormGamma} \\
        & \begin{bmatrix} -\phi && \ast \\ \mathbf{U} && -1 \end{bmatrix} \leq 0, \label{eq:minNormU} \\
        & c(\tilde{\mathbf{y}}) \leq 0, \label{eq:minNormOthers}
        \end{alignat}
\end{subequations}
where \eqref{eq:minNormOthers} encapsulates all the constraints in Instantiation\;\ref{insta:semiProg} together with any modifications or added constraints suggested by any of Theorems\;\ref{thm:cpaSafety}, \ref{thm:singleStage}, or \ref{thm:multiStage}. Note that \eqref{eq:minNormU} is the Schur complement of $\mathbf{U}^\intercal\mathbf{U} - \phi \leq 0$. Since $\underline{\mathbf{y}}$ is a feasible point of \eqref{eq:minNormSDP}, the convex-overbounding technique used in Theorem\;\ref{thm:semiProg2} can be applied to solve \;\ref{eq:minNormSDP} iteratively. This gives an example for continuing Algorithm\;\ref{alg:cpaControl} after line\;\ref{line:term1}.

\subsection{QP-Based Online Application}
As discussed, the restriction of the obtained Lyapunov-like functions to their respective positive-invariant sets in Theorems\;\ref{thm:cpaStability}, \ref{thm:cpaSafety},  \ref{thm:singleStage}, and \ref{thm:multiStage} are either \ac{CLF}s or \ac{CBF}s. Thus online \ac{QP} can be implemented to find a point-wise minimum-norm controller as the state trajectory evolves similar to what is proposed in \cite{Ames2016}.

Suppose that $\underline{b}_2>0$ is found by Algorithm\;\ref{alg:cpaControl}, and $\underline{V}:\A\rightarrow\R$ is the restriction of the corresponding CPA Lyapunov function to one of its sublevel sets inside the triangulation as specified in any of Theorems\;\ref{thm:cpaStability}--\ref{thm:multiStage}, that is $\A{=}{V^\ast}^{-1}([0,r])$, $r{>}0$, where $\A{\subseteq}\T$ and $\A{\in}\mathfrak{R}^n$. Starting at any $x{\in}\A\degree$, the minimum-norm online \ac{QP} controller can be written as 
\begin{subequations} \label{eq:minNormLips} 
    \begin{flalign} 
        &u^\ast = \argmin_{u} \;\; u^\intercal\hat{H}(x)u + \hat{h}(x)^\intercal u & \nonumber \\
        &\textrm{s.t.\;\;} Hu\leq h_c, & \\
        & \quad\;\;\, \nabla{\underline{V}_i}^\intercal(f(x){+}G(x)u){+} \underline{b}_2\, \underline{V}(x) \leq 0, \;\; \forall i\in\mathcal{I}, &
        \end{flalign}
\end{subequations}

\noindent where $\mathcal{I} {=} \Set{i{\in}\IntSet_1^{\mt} | x{\in}\sigma_i}$, and $\hat{H}(x)$ is positive definite. The set $\mathcal{I}$ has more than one element if $x$ is on the common face of some simplexes. The optimization \eqref{eq:minNormLips} is feasible for all $x\in\A$, because the corresponding CPA controller of $V^\ast$ is a feasible point for it. Therefore, the convergence inequality $||x(t)||\leq\sqrt[^{\underline{a}}]{r/\underline{b}_1}e^{-( \underline{b}_2/\underline{a})(t-t_0)}$ 
that holds for the CPA controller, also holds for the QP-based controller.
 
\section{Triangulation Refinement}\label{sc:TriangulRefinement}
Theorems\;\ref{thm:cpaStability}--\ref{thm:multiStage} and their iterative implementation via Algorithm\;\ref{alg:cpaControl} work on given, fixed triangulations. If $b_2>0$ is not found, the triangulation can be refined, introducing more simplexes in the same triangulable set. In \eqref{eq:betaAndc}, $c_{i,j}$, roughly representing the length of an edge in each simplex squared, is multiplied by $\beta_i$ which compensates for higher order terms in the Taylor's theorem. Thus, reducing the length of edges leads to tighter upper-bounds on $D^+V(\cdot)$. These triangulation refinements can be local by tracking the value of $D^+_{i,j}V$ on the simplexes in $\T$ as a numerical solver searches for $b_2>0$. Let $\Omega\subseteq\X$ be an admissible triangulable set, and $\mathcal{H}$ the set of constraint surfaces for the triangulation of $\Omega$, and $\rho:\hat{\Omega}\rightarrow\R_{>0}$, where $\hat{\Omega}\subseteq\Omega$, be a function representing simplex sizes in a region of interest. Algorithm\;\ref{alg:modification} describes a very general way of refining triangulations for Instantiation\;\ref{insta:semiProg}'s optimization or its iterative version in Algorithm\;\ref{alg:cpaControl} because Theorems\;\ref{thm:cpaStability}--\ref{thm:multiStage} and their iterative versions are formulated using those.

\begin{algorithm}[h!]
	\caption{Control design with triangulation refinement}
    \label{alg:modification}
	\begin{algorithmic}[1]
	   \Require System \eqref{eq:controlAffineSystem}, a triangulable set $\Omega\subseteq\X$, cost function, simplex size function $\rho:\hat{\Omega}\rightarrow\R_{>0}$, where $\hat{\Omega}\subseteq\Omega$, minimum simplex size $\rho_{\textrm{min}}$, constraint surfaces $\mathcal{H}$,  $0{<}\gamma{<}1$.
	   \Ensure $\mathbf{y} =  [\mathbf{V}, \mathbf{L}, \mathbf{U}, \mathbf{Z}, a, \mathbf{b}]$
	   \Repeat{}
	        \State Generate $\T$, the $\Omega$'s triangulation respecting $\rho$ and $\mathcal{H}$
	        \State Solve \eqref{eq:SemiProg} or use  Algorithm\;\ref{alg:cpaControl}
	        \If{desired objectives are met}
	            \State Return $\mathbf{y}$ \label{line:refFindingA}
	        \EndIf
	        \State $\rho := \gamma \rho$
	   \Until{$\rho_{\textrm{min}}$} is reached
	\end{algorithmic}
\end{algorithm}

\section{Numerical Examples}\label{sc:Examples}
In this section, controller design for a constrained nonlinear system using the methods introduced in this paper is conducted and compared to another method. Further, applications of the method for some special cases are considered. All computations were conducted in MATLAB\;2019b on a desktop computer with an AMD Ryzen\;5\;2600 six-core CPU, and 8\;GB\;DDR4 RAM, running the 64-bit version of Windows\;10.

To aid visualization, all examples were conducted on a 2D system that models a pendulum as a lumped mass connected to a fixed revolute joint with a mass-less rod. The equations of motion and constrains are
\begin{subequations}\label{eq:pendulum}
    \begin{align} 
    \dot{x} &= g(x,u), \\
    ||x||_{\infty} &\leq 1, \;\; |u| \leq 5,
\end{align}
\end{subequations}
where $g(x,u)=[x^{(2)}\;\;4.9\sin{x^{(1)}} {-} 0.3x^{(2)} {+} u]^\intercal$, and all units are SI. The point $(x,u)=(0,0)$ denotes the unstable, unforced equilibrium of \eqref{eq:pendulum} corresponding to the pendulum at its upright position with zero angular velocity.

\subsection{Controllers}
The design methods for \eqref{eq:pendulum} are as follows.

\subsubsection{Dynamic Programming (DP)}
Probably the closest design method to that of this paper in terms of its generality, independency to \textit{a-priori} design choices, respecting constraints, and offering a complete offline design method by minimizing a cost function, is the \ac{DP} solution of a finite-time optimal control problem. Moreover, solving DP by starting at a control-invariant set and going backward in time, makes it similar to the multi-stage design of this paper since stabilizable sets are known to be contained in their precursor sets \cite[Ch\;10]{BorrelliBook}.  In practice, one needs to grid the state-space and discretize the equations of motion to be able to compute stabilizable sets and evaluate the value function by interpolation, thereby losing accuracy in finding the boundaries of the stabilizable sets and optimality of the solutions. Having approximate boundaries may result in having a stabilizable set that is not contained in its precursor set. Moreover, finding a suitable sampling-time and grid size to balance computation time and accuracy are non-trivial. 

Assuming a finite horizon of $N$ steps, and $\X_{N\rightarrow N}$ a control-invariant terminal set, $\X_{k\rightarrow N}$ denotes the $(N{-}k)$-step stabilizable set defined as the set of feasibles states that can be driven to $\X_{N\rightarrow N}$ in $k$ steps using a sequence of $k$ admissible inputs \cite[Ch\;10]{BorrelliBook}. Having $\{\X_{k\rightarrow N}\}_{k=0}^N$, the following optimization is solved for $k\in\IntSet_{k=0}^{N-1}$ backwards for $\forall x\in\X_{k\rightarrow N}$ to determine the time-varying \ac{DP} controller.

\begin{subequations} \label{eq:DP} 
    \begin{alignat}{2}
        &J^\ast_{k\rightarrow N}(x) = \min_{u(x)} \;\; x^\intercal x + J^\ast_{k+1\rightarrow N} (x^+) \nonumber \\
        \textrm{s.t.} \;\; & x^+ = x + T_s\; g(x,u(x)), \label{eq:descretizedEOM}\\
        & x^+ \in \X, \;\; u(x)\in\U, \\
        & x^+ \in \X_{k+1 \rightarrow N},
        \end{alignat}
\end{subequations}
where $J^\ast_{N\rightarrow N}(x)=x^\intercal x$, and $J^\ast_{k\rightarrow N}(\cdot)$ is the optimal cost-to-go of the step $k$, and \eqref{eq:descretizedEOM} denotes the Euler discretization of \eqref{eq:pendulum}, and $T_s$ is the sampling time, and $\X_{k+1\rightarrow N}$ is the $(N{-}k{-}1)$-step stabilizable set. The quadratic cost function in \eqref{eq:DP} selects controllers that result in fast convergence by not penalizing $u(x)$, making it competitive to CPA designs of this paper. To compute approximations of $\{\X_{k\rightarrow N}\}_{k=0}^{N-1}$ and $\{J^\ast_{k\rightarrow N}(\cdot)\}_{k=0}^{N-1}$, a uniform grid for the state-space with step size $\delta x^{(1)} = \delta x^{(2)}$ is assumed. Since $\X_{N\rightarrow N}$ is given, each element of $\{\X_{k\rightarrow N}\}_{k=0}^{N-1}$ is approximated as the convex hull of all the grid points $x$ at which \eqref{eq:DP} is feasible. Thus at each grid point, a feasibility problem with linear constraints is formulated for each $k\in\IntSet_{k=0}^{N-1}$. Then, \eqref{eq:DP} is solved backwards at the grid points inside the stabilizable sets given $J^\ast_{N\rightarrow N}(x)$ and using interpolation to evaluate optimal cost-to-goes.

The uniform grid $\delta x^{(1)}=\delta x^{(2)} = 0.02$ was chosen. To minimize the number of \textit{a-priori} design choices, the origin was selected as the terminal set in \eqref{eq:DP}. However, due to inevitable inaccuracies in finding $k$-step stabilizable sets, $||x||_\infty\leq 0.01$ was selected as $\X_{N\rightarrow N}$ instead of $\X_{N\rightarrow N}=0$ to avoid numerical issues. In order to have a four-second simulation time, $T_s = 0.1$ s and $N = 40$ were chosen for \eqref{eq:DP}. Using this setup and the resulting controller, any $x\in\X_{0\rightarrow 40}$ is expected to reach $||x||_\infty\leq0.01$ in four seconds while satisfying state and input constraints, but inaccuracies  result in slight violations. For instance,
\begin{equation}\label{eq:containedness}
    \X_{k+1\rightarrow 40}\subseteq\X_{k\rightarrow 40}
\end{equation}
hold for all $k\in\IntSet_{0}^{40}$ except for $k\in\{28,12,4,0\}$. These choices provided an acceptable balance between accuracy and computation time. That is, coarser grid sizes made violations of \eqref{eq:containedness} more frequent, while finer ones increased computation time. Stabilizable sets were computed by solving the discussed linear feasibility problems in SeDuMi \cite{sedumi}. Then, \eqref{eq:DP} was solved at each grid point in the corresponding stabilizable set for $k\in\IntSet_0^{N-1}$ using the nonlinear optimization solver `fmincon' by assuming `spline' interpolation to evaluate the optimal cost-to-goes.

\subsubsection{CPA controllers}
Two CPA controllers were designed. The software package Mesh2d \cite{mesh2d,mesh2dPack} was used to generate triangulations, in which modifying the maximum edge size function $\rho:\R^n\rightarrow\R_{>0}$ caused refinements. All \ac{SDP}s were solved by SeDuMi. The refinements were implemented as characterized by Algorithm\;\ref{alg:modification}, respecting constraint surfaces. For \ac{SDP} initializations, LQR with the cost function $2x^Tx + u^2$ was used. 

\begin{figure}
	\centering
	\begin{subfigure}[t]{0.45\linewidth}
		\centering
		\includegraphics[width=2.6cm]{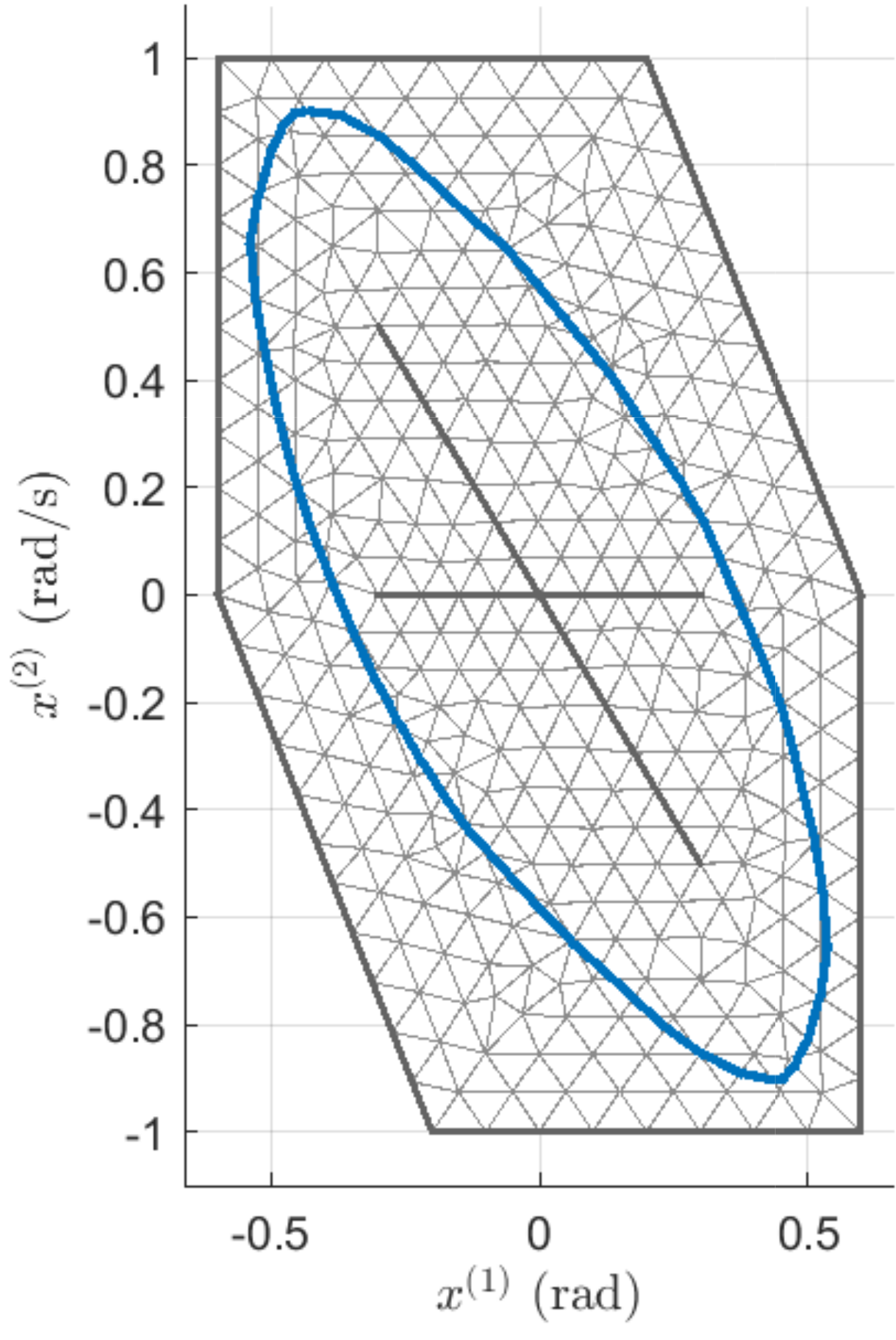}
		\caption{First stage}\label{fig:conEdges1}
	\end{subfigure}
	\quad
	\begin{subfigure}[t]{0.45\linewidth}
		\centering
		\includegraphics[width=4cm]{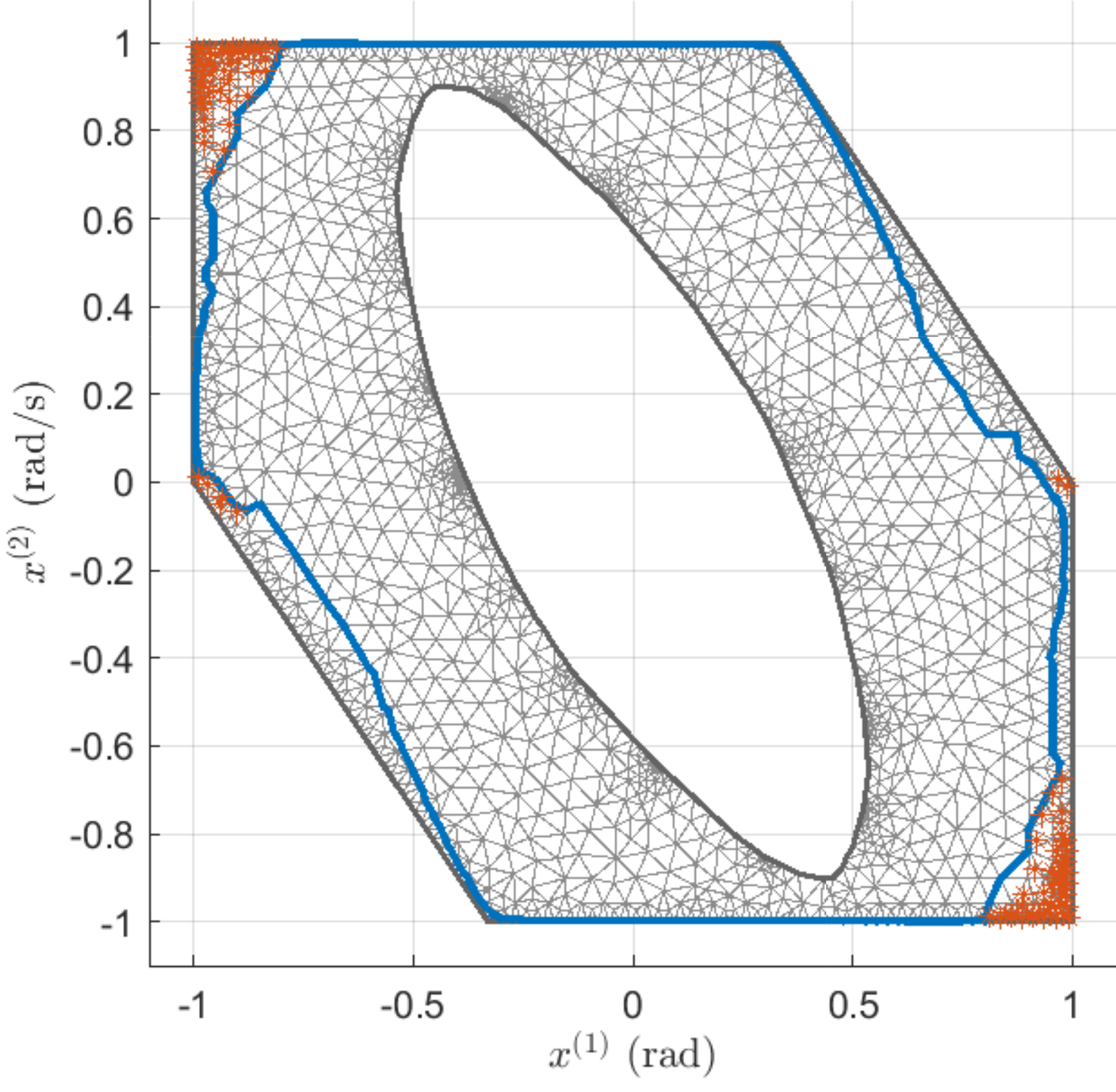}
		\caption{Second stage}\label{fig:fistStageSet1}
	\end{subfigure}
	\\
	\begin{subfigure}[t]{0.45\linewidth}
		\centering
		\includegraphics[width=4cm]{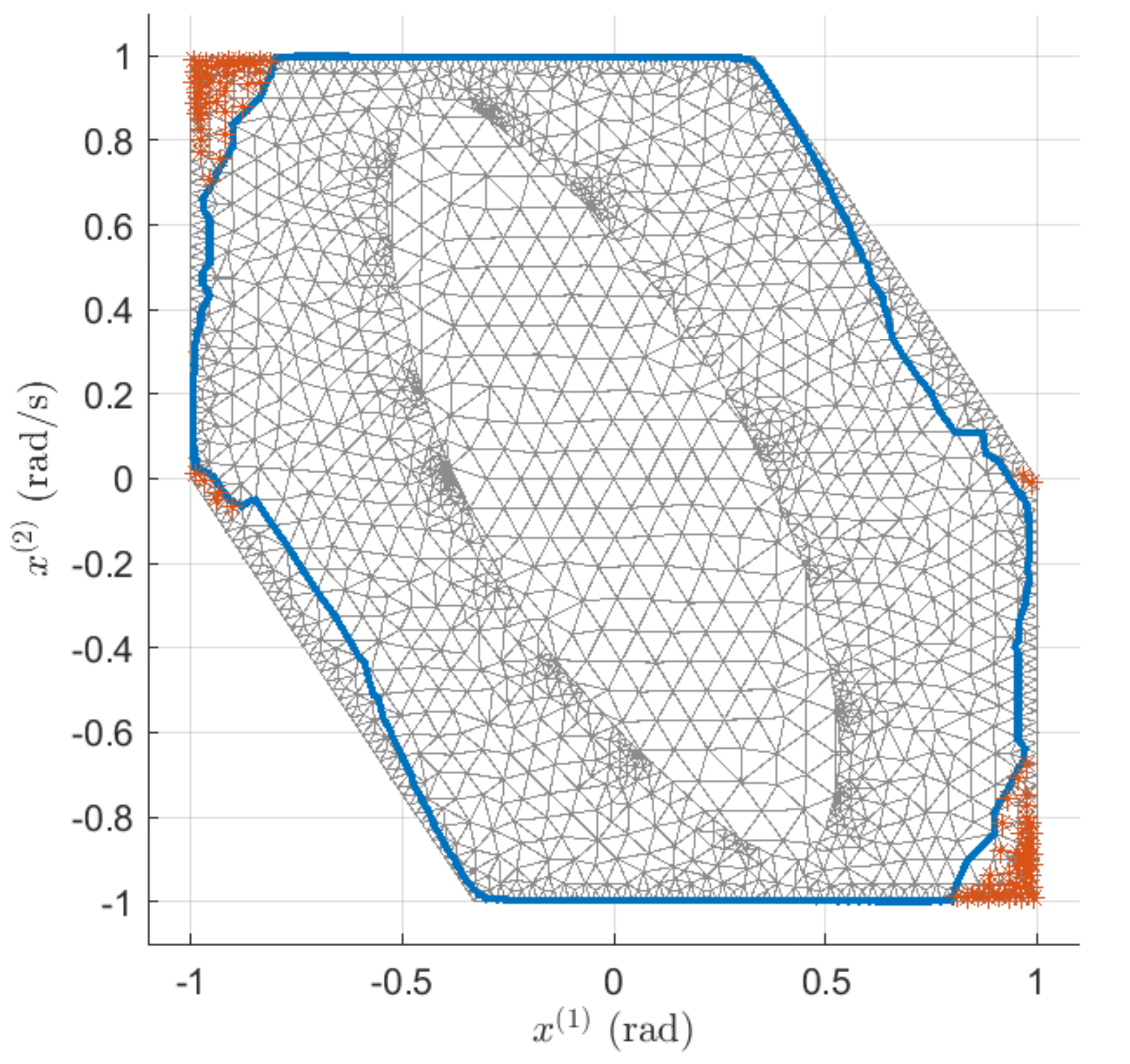}
		\caption{Combined (Top view)}\label{fig:conEdges2}
	\end{subfigure}
	\quad
	\begin{subfigure}[t]{0.45\linewidth}
		\centering
		\includegraphics[width=4cm]{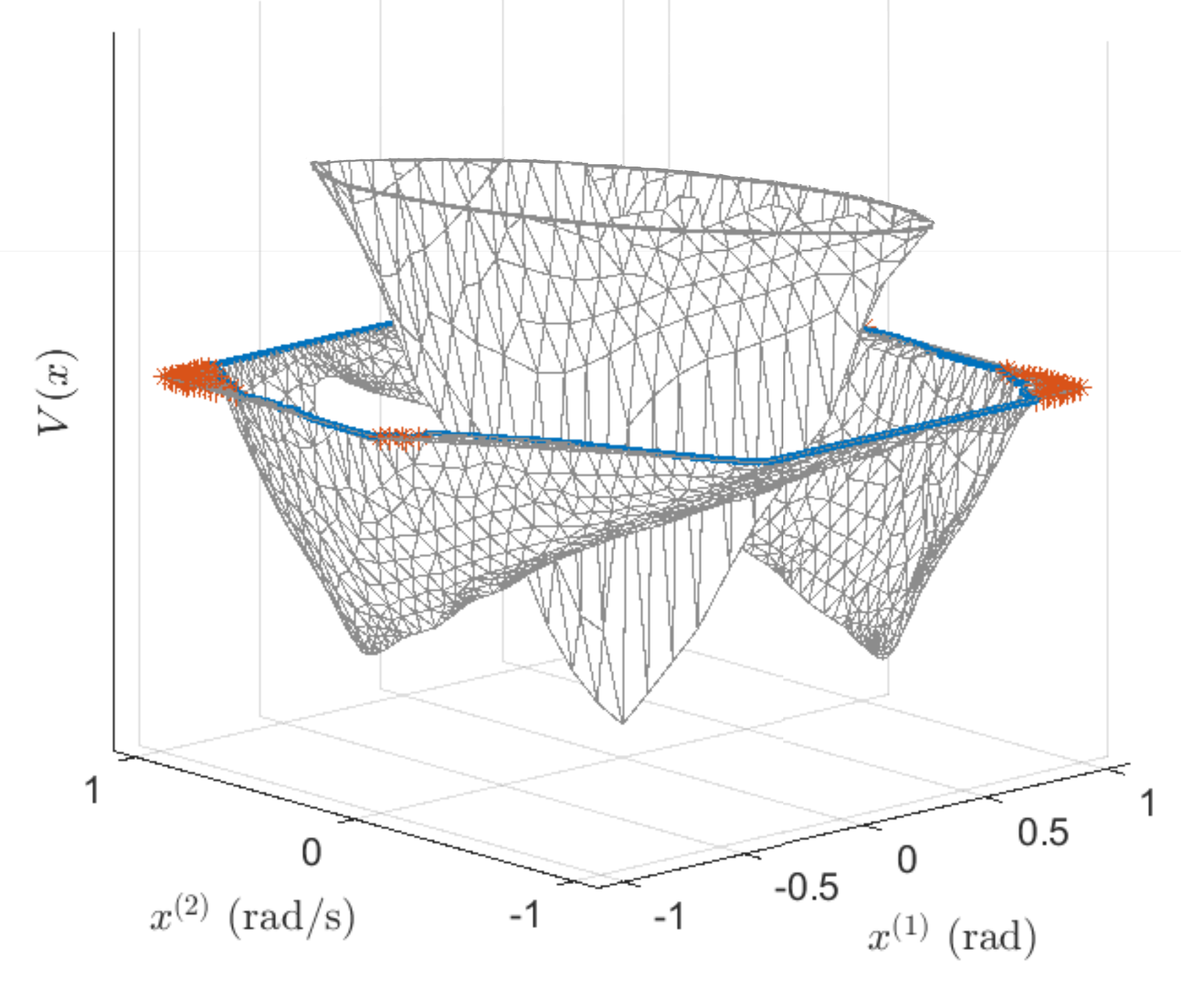}
		\caption{Combined (3D view)}\label{fig:fistStageSet2}
	\end{subfigure}
	\caption{Triangulations and the corresponding \ac{CPA} Lyapunov-like functions in the two-stage design. The thick gray lines in (a) and (b) are constraint surfaces, and the blue lines represent the level sets of the Lyapunov-like functions. The level set found in the first stage is the inner constraint surface in the second. Combined, the discontinuous CPA Lyapunov function and its level set is depicted in (c) and (d).  The simplexes marked by asterisks have $b_2\leq0$ in one or more vertices.}\label{fig:multiStage}
    \end{figure}

The second stage was designed using Theorem\;\ref{thm:multiStage} and Remark\;\ref{rem:discontVu}, where both the Lyapunov function and controller were allowed to be discontinuous along the level set of the first stage. The constraint surfaces of the second stage are depicted in Fig.\;\ref{fig:multiStage}(b) with thick gray lines. The inner one is the level set found in the first stage, and the boundary is again a very crude polygon resembling the slanted ecliptic level sets of the LQR initialization. For the initial triangulation and its refinements, $\rho(\cdot)=0.1$ was used everywhere, while $\gamma = 0.5$ and $\gamma=0.8$ were used for the boundary and elsewhere, respectively. Note that the simplex sizes around the inner boundary are imposed by the level set found in the first stage. Due to the presence of a large number of vertices that form the level set, the simplex sizes are automatically adjusted in their vicinity by Mesh2d to generate a triangulation. The iterative algorithm was allowed to complete 10 iterations on each triangulation. Since the level sets found after two sets of 10 iterations were still small, another refinement was performed, making $\rho(\cdot)=0.5^2\times0.1$ and $0.8^2\times0.1$ on the boundary and elsewhere, respectively. The corresponding triangulation, and the level set found at the tenth iteration on this triangulation are also given in Fig.\ref{fig:multiStage}(b). This level set was found by Corollary\;\ref{cor:mightFindSafetyController} since a positive $b_2$ was not found. The simplexes marked with red asterisks are the ones that have $D^+V\leq0$ in at least one of their vertices. Although the triangulation and \eqref{eq:pendulum} are symmetric with respect to the origin, the resulting level set of the Lyapunov-like function is not due to numerical errors in the \ac{SDP} solver. The solution was checked in the original optimization from which the \ac{SDP} was formulated to make sure it was a feasible point. The combined triangulation from the two stages and the resulting discontinuous \ac{CPA} Lyapunov function are depicted in Fig.\;\ref{fig:multiStage}(c) and (d), respectively.

\begin{figure}
	\centering
	\begin{subfigure}[t]{0.45\linewidth}
		\centering
		\includegraphics[width=3.5cm]{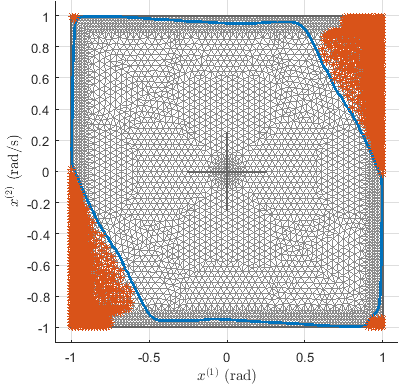}
		\caption{Single-stage design}\label{fig:conEdges}
	\end{subfigure}
	\quad
	\begin{subfigure}[t]{0.45\linewidth}
		\centering
		\includegraphics[width=3.5cm]{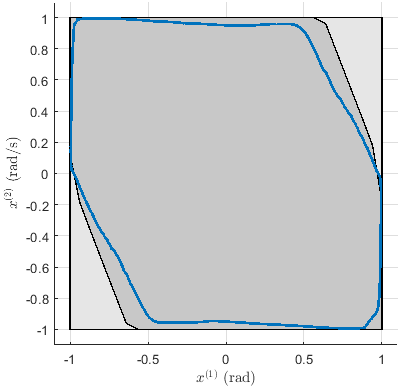}
		\caption{Comparing the \ac{ROA}}\label{fig:fistStageSet}
	\end{subfigure}
	\caption{Single-stage design and comparison with 40-step \ac{DP}. In (a), thick gray lines represent the constraint surfaces of the triangulation, and the simplexes marked by asterisks have $b_2\leq0$ in one or more vertices. In (b), light and dark gray areas are $\X$ and the 40-step stabilizable sets in \ac{DP}, respectively. The blue line represents a level set of the CPA Lyapunov function, that is identical in (a) and (b). Although the ROA of the single-stage design is 5\% smaller than that of 40-step DP, it was computed in 28\% of the time.}\label{fig:myROAfig}
    \end{figure}

\textbf{2.2)} Single-stage design: Here, Theorem\;\ref{thm:singleStage} and Corollary\;\ref{cor:mightFindSingleStage} were used. Instead of choosing a custom set for triangulation, the whole $\X$ in \eqref{eq:pendulum} was triangulated to let the algorithm maximize the \ac{ROA}. Constraint surfaces for this design were the boundary of the triangulation and the two surfaces passing through the origin. They are depicted in Fig.\ref{fig:myROAfig}(a) with thick gray lines. They are the boundary of $\X$, and two vertical lines passing through the origin to make Mesh2d generate identical simplexes around the origin. For simplex sizes, $\rho(\cdot)=0.02$ was used in $||x||\leq0.1$ and $0.9\leq||x||_\infty\leq 1$ regions, while $\rho(\cdot)=\rho=0.04$ was assigned to elsewhere.  After 8 iterations, the level set depicted in Fig.\ref{fig:myROAfig}(a) was found.

\subsection{Comparison}
Qualitatively, since the time-varying \ac{DP} controller is obtained by solving \eqref{eq:DP} only on grid points, it has a suboptimal performance. Moreover, the stabilizable set boundaries are approximate. This can lead to slight violations of the state or input constraints. On the other hand, the \ac{CPA} controllers are conservative since they bound the maximum element of the closed-loop system's Hessian above, and they are obtained by convex-overbounding. However, once a solution is found in them, the boundary of the \ac{ROA} is exact and no constraint violations are possible once the closed-loop system is initialized inside it. The \ac{DP} and \ac{CPA} controllers were compared based on the offline synthesis time and the closed-loop settling time to $||x||\leq 0.01$, starting at identical points.

\subsubsection{Synthesis time}
For CPA controllers, the synthesis time is the time required to solve all the \ac{SDP} iterations, including the time spent on coarser triangulations that did not yield the final solution. However, for \ac{DP}, the synthesis time only involves the computation time on the discussed grid, thus omitting the time spent on the trial-and-error process on both finer and coarser grids. Therefore, this synthesis time comparison gave an advantage to \ac{DP}. 

\begin{figure}
	\centering
	\begin{subfigure}[t]{0.45\linewidth}
		\centering
		\includegraphics[width=4cm]{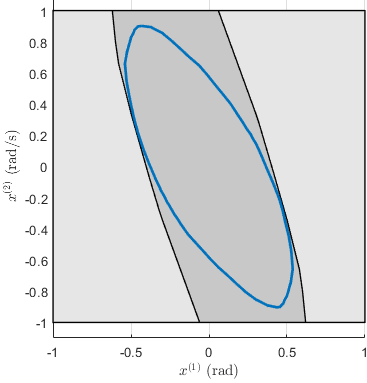}
		\caption{First stage}\label{fig:conEdges3}
	\end{subfigure}
	\quad
	\begin{subfigure}[t]{0.45\linewidth}
		\centering
		\includegraphics[width=4cm]{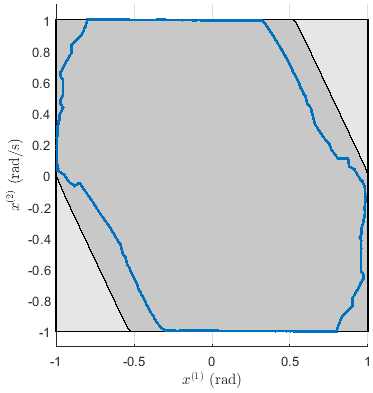}
		\caption{Second stage}\label{fig:fistStageSet3}
	\end{subfigure}
	\caption{Comparing the level sets of the multi-stage design with stabilizable sets of \ac{DP}. The light gray area represents $\X$. The dark gray areas represent 8-step stabilizable set in (a), and 18-step stabilizable set in (b). The level sets from the first and second stages of the two-stage design, depicted in blue, are inside the 8-step and 18-step stabilizer sets of the origin in \ac{DP}, respectively. Although the ROA of the combined stages is 18\% smaller than that of 18-step DP, it was computed in 26\% of the time.}\label{fig:multiStageCOMPARE}
    \end{figure}

The level set found in the first stage of the two-stage design is inside 8-step stabilizable set found by \ac{DP}, as depicted in Fig.\;\ref{fig:multiStageCOMPARE}(a). The level set found in the second stage of the two-stage design is inside 18-step stabilizable set found by \ac{DP}, as depicted in Fig.\;\ref{fig:multiStageCOMPARE}(b). The syntheis times and the areas of the \ac{ROA}s are given in Table\;\ref{tab:results}. It shows that comparable \ac{ROA}'s are found significantly faster for \ac{CPA} controllers. Note that in DP, the area of the 18-step and 40-step stabilizable sets are almost the same, meaning that during the intermediate steps, the DP cost is decreasing while the ROA is not growing. It is worth mentioning that during 100.8\;min, which is almost equal to the time spent for the single-stage CPA design, the \ac{DP} had finished only 13-steps.

\begin{table}[]
\centering
\caption{Syntheis time and area of ROA comparison}
\begin{tabular}{lcc}
\multicolumn{1}{c|}{Controller}                    & \multicolumn{1}{c|}{Synthesis time (min)} & ROA's area   \\ \hline
\multicolumn{1}{l|}{Two-stage CPA (first stage)}   & \multicolumn{1}{c|}{13.87}                & 0.27         \\
\multicolumn{1}{l|}{DP (8 steps)}                  & \multicolumn{1}{c|}{54.90}                & 0.37         \\ \hline
\multicolumn{1}{l|}{Two-stage CPA (second stage)}  & \multicolumn{1}{c|}{27.09}                & 0.45$^{\ast\;\,}$ \\
\multicolumn{1}{l|}{DP (the next 10 steps)}        & \multicolumn{1}{c|}{101.92}               & 0.51$^{\ast\ast}$ \\ \hline
\multicolumn{1}{l|}{Two-stage CPA (Combined)}      & \multicolumn{1}{c|}{40.96}                & 0.72         \\
\multicolumn{1}{l|}{DP (18 steps)}                 & \multicolumn{1}{c|}{156.82}               & 0.88         \\ \hline
\multicolumn{1}{l|}{Single-stage CPA}              & \multicolumn{1}{c|}{97.20}                & 0.86         \\
\multicolumn{1}{l|}{DP (40 steps)}                 & \multicolumn{1}{c|}{351.6}                & 0.91         \\ \hline
\multicolumn{3}{l}{$^{\ast\;\,}$ \footnotesize area of the hollowed region}                     \\
\multicolumn{3}{l}{$^{\ast\ast}$ \footnotesize area of 18-step stabilizable set minus 8-step's}
\end{tabular} \label{tab:results}
\end{table}



\subsection{Settling time}
\begin{figure}
       \centering
       \includegraphics[width=5cm]{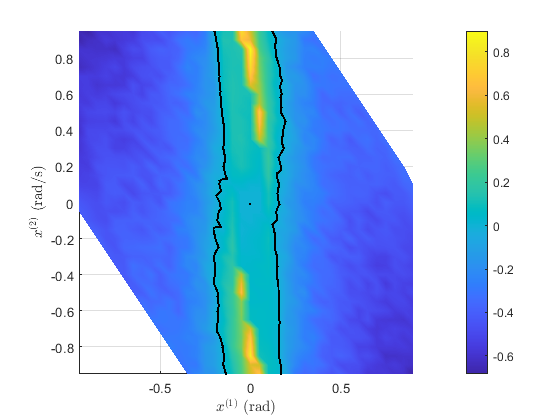}
       \caption{The difference between the settling time, in seconds, of the state norm to the ball of radius 0.01 using the \ac{DP} controller and the two-stage design, initialized at identical points. The black line represents the zero level set. The two-stage CPA design settles faster for most of the initial points in its \ac{ROA}.}
       \label{fig:deltaSettle}
   \end{figure}

The settling time of the closed-loop system using the two-stage design and \ac{DP} to $||x||\leq0.01$ was compared by initializing them on a uniform grid with $\delta x^{(1)}=\delta x^{(2)} = 0.05$ inside the \ac{ROA} of the two-stage controller. In Fig.\;\ref{fig:deltaSettle}, the difference in the settling time is given, showing that the closed-loop system with the two-stage \ac{CPA} design converges faster than \ac{DP} in most regions of the \ac{ROA}.



\section{Conclusion}
In this paper, a systematic control synthesis method for state- and input-constrained nonlinear systems was developed using \ac{CPA} Lyapunov functions and controllers on triangulated subsets of the admissible states. The method is distinguished by its generality, complete offline design, and independence from typical unclear \textit{a priori} design choices. For control-affine systems, the method was formulated as efficient iterative \ac{SDP}s that can be solved using available software. Further, a minimum-norm controller was introduced. Safety and stability were guaranteed by finding \ac{CBF} or \ac{CLF}s, and the controller simultaneously. Therefore, it can be also viewed as a systematic approach to find Lipschitz \ac{CBF}s and \ac{CLF}s. Numerical examples showed the efficiency and effectiveness of the method. The future work includes extensions of the method to switched and uncertain systems.

\bibliographystyle{unsrt}  
\bibliography{BibliographyShort.bib}
\begin{acronym}
\acro{SDP}{semi-definite program}
\acro{MPC}{model predictive control}
\acro{CLF}{control Lyapunov function}
\acro{CBF}{control barrier function}
\acro{CPA}{continuous piecewise affine}
\acro{QP}{quadratic programming}
\acro{DP}{dynamic programming}
\acro{ROA}{region of attraction}
\end{acronym}



\end{document}